\documentclass[12pt]{article}
\pdfoutput=1

\usepackage{graphics,amssymb,float}
\usepackage{graphicx}
\usepackage{rotating}
\usepackage{dcolumn}
\usepackage{bm}
\usepackage{cite}
\usepackage{amsmath}
\usepackage{indentfirst} 
\usepackage{epstopdf}

\def\CU{C_U}
\def\CD{C_D}
\def\CL{C_L}
\def\CV{C_V}
\def\CG{C_g}
\def\CP{C_\gamma}
\def\cu{\CU}
\def\cd{\CD}
\def\cv{\CV}
\def\cg{\CG}
\def\cp{\CP}

\def\tanb{\tan\beta}

\def\chimin{\chi^2_{\rm min}}

\def\gam{\gamma}
\def\ie{{\it i.e.}}
\def\eg{{\it e.g.}}
\def\what{\widehat}
\def\dcg{\Delta \CG}
\def\dcp{\Delta \CP}

\def\cl{\CL}
\def\gev{~{\rm GeV}}

\def\chimin{\chi^2_{\rm min}}
\def\anti{\overline}
\def\fbi{~{\rm fb}^{-1}}
\def\tev{~{\rm TeV}}
\def\bit{\begin{itemize}}
\def\eit{\end{itemize}}
\def\ben{\begin{enumerate}}
\def\een{\end{enumerate}}
\def\beq{\begin{equation}}
\def\eeq{\end{equation}}

\def\brinv{{\cal B}(H\to {\rm invisible})}
\def\br{{\cal B}}
\def\rts{\sqrt{s}}
\def\rinv{R_{\rm inv}}

\def\lsim{\mathrel{\raise.3ex\hbox{$<$\kern-.75em\lower1ex\hbox{$\sim$}}}}
\def\gsim{\mathrel{\raise.3ex\hbox{$>$\kern-.75em\lower1ex\hbox{$\sim$}}}}
\def\ifmath#1{\relax\ifmmode #1\else $#1$\fi}

\begin{document}
\begin{titlepage}
\begin{center}

\vspace*{-1cm}
\begin{flushright}
LAPTH-013/13\\
LPSC 13039\\
LPT 12-53\\
UCD-2013-2
\end{flushright}

\vspace*{2cm}
{\Large\bf Status of invisible Higgs decays} 

\vspace*{1cm}\renewcommand{\thefootnote}{\fnsymbol{footnote}}

{\large 
G.~B\'elanger$^{1}$\footnote[1]{Email: belanger@lapp.in2p3.fr},
B.~Dumont$^{2}$\footnote[2]{Email: dumont@lpsc.in2p3.fr},
U.~Ellwanger$^{3}$\footnote[3]{Email: ulrich.ellwanger@th.u-psud.fr},
J.~F.~Gunion$^{4,5}$\footnote[4]{Email: jfgunion@ucdavis.edu}, 
S.~Kraml$^{2}$\footnote[5]{Email: sabine.kraml@lpsc.in2p3.fr}
} 

\renewcommand{\thefootnote}{\arabic{footnote}}

\vspace*{1cm} 
{\normalsize \it 
$^1\,$LAPTH, Universit\'e de Savoie, CNRS, B.P.110, F-74941 Annecy-le-Vieux Cedex, France\\[1mm]
$^2\,$Laboratoire de Physique Subatomique et de Cosmologie, UJF Grenoble 1,
CNRS/IN2P3, INPG, 53 Avenue des Martyrs, F-38026 Grenoble, France\\[1mm]
$^3\,$Laboratoire de Physique Th\'eorique, UMR 8627, CNRS and
Universit\'e de Paris--Sud, F-91405 Orsay, France\\[1mm]
$^4\,$Department of Physics, University of California, Davis, CA 95616, USA\\[1mm]
$^5\,$Kavli Institute for Theoretical Physics, University of California, Santa Barbara,\\ CA 93106-4030, USA}

\vspace{1cm}

\begin{abstract}
We analyze the extent to which the LHC and Tevatron results as of the end of 2012 constrain 
invisible (or undetected) decays of the Higgs boson-like state at $\sim 125\gev$. To this end 
we perform global fits for several cases: 
1)~a Higgs boson with Standard Model (SM) couplings 
but additional invisible decay modes; 
2)~SM couplings to fermions and vector bosons, but allowing for additional new particles modifying the effective Higgs couplings to gluons and photons; 
3)~no new particles in the loops but tree-level Higgs couplings to the up-quarks, down-quarks and
vector bosons, relative to the SM, treated as free parameters. 
We find that in the three cases invisible decay rates of 23\%, 61\%, 88\%, respectively, 
are consistent with current data at 95\% confidence level (CL).
Limiting the coupling to vector bosons, $\cv$, to $\cv \le 1$ in case 3) reduces the allowed invisible branching ratio 
to 56\% at 95\% CL. Requiring in addition that the Higgs couplings to quarks have the same sign as in the SM,  
an invisible rate of up to 36\% is allowed at 95\%~CL. 
We also discuss direct probes of invisible Higgs decays, as well as the interplay with dark matter searches. 
\end{abstract}

\end{center}

\end{titlepage}

\section{Introduction}

The recent discovery~\cite{atlas:2012gk, cms:2012gu} of a new particle with properties consistent with a Standard Model (SM) Higgs boson is clearly the most significant news from the Large Hadron Collider (LHC). 
This discovery was supported by evidence for a Higgs boson found by the CDF and D0 collaborations at the
Tevatron \cite{tevatron:2012zzl} and completes our picture of the SM. 
However, the SM leaves many fundamental questions open---perhaps the most pressing issue being that the SM does not explain the value of the electroweak scale, \ie\  the Higgs mass, itself.  Clearly, a prime goal after the discovery is to thoroughly test the SM nature of this Higgs-like signal. 

With the measurements in various channels, a comprehensive study of the properties of the Higgs-like state becomes possible and has the potential for revealing whether or not the Higgs sector is as simple as envisioned in the SM.  
In a recent study~\cite{Belanger:2012gc}, we analyzed the extent
to which the results from the LHC and Tevatron, as published by the end of 2012, constrain the
couplings of the Higgs boson-like state at $\sim 125\gev$. To this end
we assumed that only SM particles appear in the Higgs
decays, but tree-level Higgs couplings to the up-quarks, down-quarks and
vector bosons, relative to the SM are free parameters. 
Moreover, we considered the case that new particles appearing in loops modify the 
effective Higgs couplings to gluons and/or photons. 
We found that the SM expectation is more than
$2\sigma$ away from fits in which: a) there is some non-SM contribution
to the $\gam\gam$  coupling of the Higgs; or b) the sign of the top
quark coupling to the Higgs is opposite that of the SM. In both
these cases good fits with $p$-values $\sim 0.9$ can be achieved. 
Since option b) is difficult to realize in realistic models, it would seem that new physics
contributions to the effective couplings of the Higgs are preferred.

In this paper we carry our work  a step further and investigate the extent to which current data constrain 
invisible decays (\eg\ $H\to \tilde \chi^0_1 \tilde \chi^0_1$, where $\tilde \chi^0_1$ is the lightest SUSY particle) 
or undetected decays (such as $H\to AA$, where $A$ is a light CP-odd, perhaps singlet scalar) 
of the Higgs-like particle.\footnote{Strictly speaking,  invisible Higgs decays are those which leave no 
traces in the detectors, \ie\ excluding 
decays into non-SM particles which are missed by the current searches. Such invisible Higgs decays appear 
in models where the Higgs boson can decay into stable neutral particles, including dark matter (DM) candidates. 
In the SM, invisible Higgs decays originate from  $H \rightarrow Z\,Z^{(*)}\rightarrow 4\ \text{neutrinos}$ with a 
small ${\cal B}(H_{\text{SM}}\to\mbox{invisible})$ of about $5.3\times 10^{-3}$. 
For the global fits, there is no difference between invisible and yet undetected decays, and we will use 
the term ``invisible" for both, genuine invisible and merely undetected. 
When talking about specific signatures, like monojets + missing energy, it is however important to make the distinction---in this case we will use ``invisible'' in the strict sense.} 

Our parametrization and fitting procedure is the same as in \cite{Belanger:2012gc}, where we introduced scaling factors $C_I$ relative to SM couplings. 
We treat the couplings to up-type and down-type fermions, $\cu$ and $\cd$, as independent parameters (but we  only consider the case $\cl=\cd$, and we assume that the $C_F$ are family universal). 
Moreover, we assume a custodial symmetry in employing a single $C_W=C_Z\equiv\CV$. 
In addition to the tree-level couplings given above, the $H$ has couplings to $gg$ and $\gam\gam$ that are first induced at one loop and are completely computable in terms of $\cu$, $\cd$ and $\cv$ if only loops containing SM particles are present. We define $\anti \cg$ and $\anti \cp$ to be the ratio of these couplings so computed to the  SM (\ie\ $\cu=\cd=\cv=1$) values.
However, in some of our fits we will also allow for additional loop contributions $\dcg$ and $\dcp$ from new particles; in this case $\cg=\anti \cg+\dcg$ and $\cp=\anti \cp +\dcp$. 

Limits on ``invisible'' Higgs decays from global fits to LHC data were obtained previously in~\cite{Espinosa:2012vu,Carmi:2012in,Giardino:2012dp,Giardino:2012ww} and within the MSSM in~\cite{Desai:2012qy}. 
Since then, much more data has become available. We therefore find it worthwhile to re-investigate the status of invisible decays. Moreover, we consider more general deviations from SM couplings than the previous works.  

The possibility  of  probing  directly the (genuine) invisible branching fraction of the Higgs in various channels at the LHC has been investigated  already some time ago~\cite{Gunion:1993jf,Choudhury:1993hv,Eboli:2000ze,Godbole:2003it,Davoudiasl:2004aj}.\footnote{Invisible Higgs decays were in fact already considered in the 1980's in \cite{Shrock:1982kd}.}
It was revisited more recently in light of the 125~GeV Higgs signal at the LHC in~\cite{Bai:2011wz,Djouadi:2012zc,Ghosh:2012ep}. The channels considered were monojets or 2 jets plus missing energy.\footnote{Ref.~\cite{Bai:2011wz} also investigated $ZH$ associated production,  leading to high-$p_T$ $Z$'s 
plus missing energy. However, stronger limits were obtained from monojet searches.} 
These  make it possible to constrain the  product of the Higgs production cross section in the gluon--gluon fusion (ggF), or 
vector boson fusion (VBF) and VH associated production modes, times the invisible branching fraction. 
Specifically, they put constraints on $\rinv({\rm ggF})= C_g^2\, \brinv$ and 
$\rinv({\rm VBF})= C_V^2\, \brinv$.
We will not include these direct limits in our fits but will comment on the impact of these probes on the valid parameter space. 
Probing yet undetected $H\to AA$ decays (with the pseudoscalars further decaying in various channels) was discussed in~\cite{Almarashi:2011te,Englert:2012wf} --- results depend strongly on the manner in which the $A$ decays.

The results of our global fits are presented in Section~2. 
In Section~3, we discuss further probes of invisible or undetected decays. 
In the case of Higgs decays into DM particles, a strong interplay with direct DM searches arises. 
This is discussed in Section~4. A summary and conclusions are given in Section~5. 

\section{Results from global fits}

\subsection*{Standard Model plus invisible decays}

\begin{figure}[t]\centering
\includegraphics[scale=0.46]{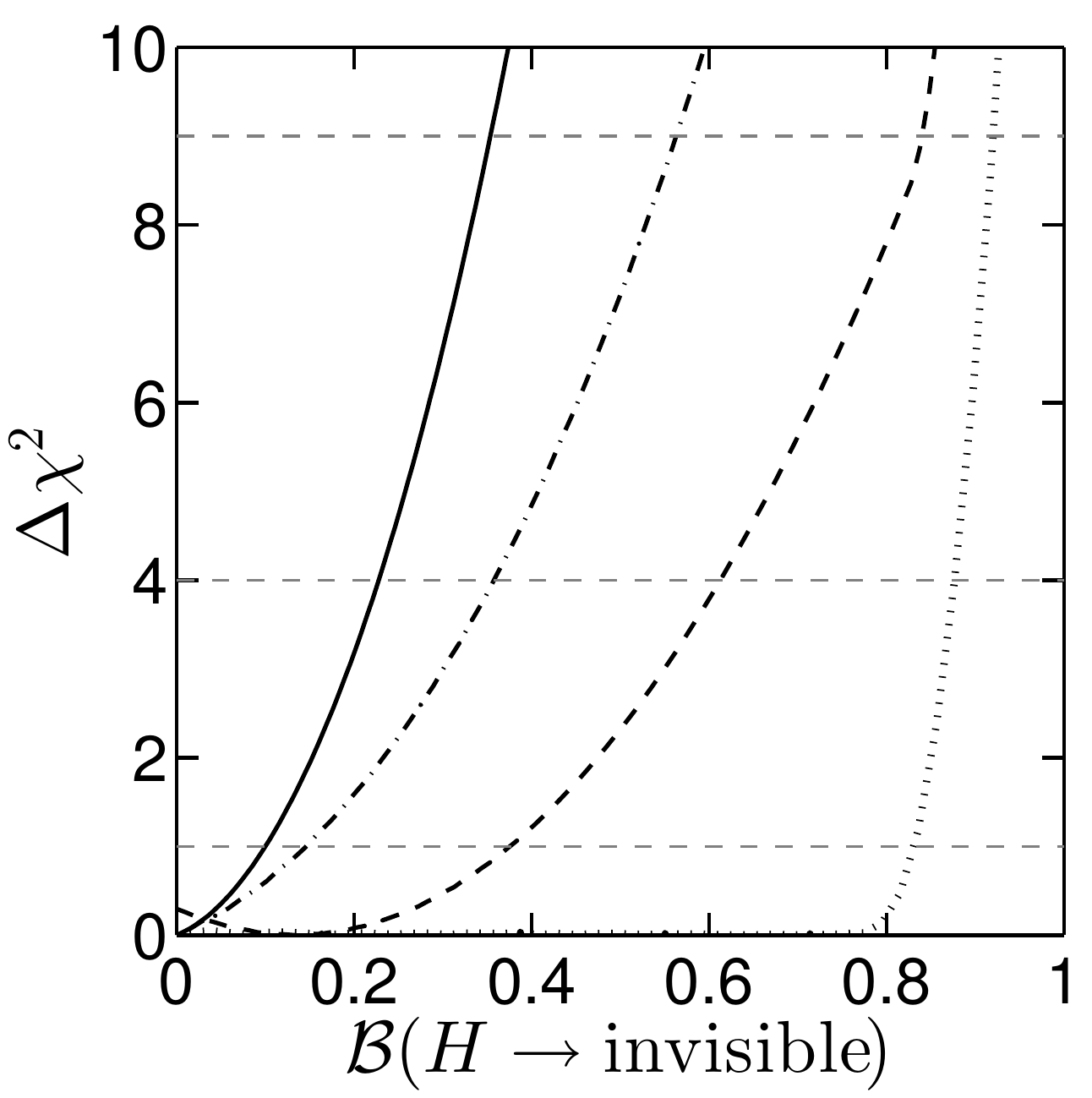}
\caption{$\Delta\chi^2$ distributions for the branching ratio of invisible Higgs decays. 
The full, dashed, and dotted lines correspond, respectively, to the cases of 
1) SM couplings, 2) arbitrary $\dcg$ and $\dcp$, and 3)~deviations of $\cu,\cd,\cv$ from unity. 
In addition, we show as dash-dotted line the variant of case 3) with $\cu,\cd>0$ and $\cv\le 1$. 
\label{fig:onedim} }
\end{figure}
 
As mentioned, we consider several cases in our analysis. We start the discussion with the simplest case, case 1),
of an SM Higgs boson augmented by invisible/undetected decays, 
\ie\ we require that all couplings be equal to the SM values ($\cu,\cd,\cv,\cg,\cp=1$) but allow for (yet) invisible decays of the Higgs state in addition to the normal decays to SM particles.  We first recall that, for the most part, the observed rates in many production-decay channels are not far from the SM-predictions and, in particular, are not suppressed.  In addition, there are a few final states that have a somewhat enhanced rate relative to the SM, most notably the $\gam\gam$ final state. Since invisible decays reduce the branching ratio to the (visible) SM final states, it is to be expected that $\brinv$ is strongly limited.  
This is confirmed in our fit. We find that $\brinv<0.23$ ($0.35$) at $2\sigma$ ($3\sigma$). 
The $\chi^2$ distribution as function of the invisible rate for case 1) is shown as the solid line in Fig.~\ref{fig:onedim}.
The $\chi^2$ at the minimum is $\chimin=20.2$, \ie\ the same as for the SM~\cite{Belanger:2012gc}. 
With 20 degrees of freedom (d.o.f.) this corresponds to a $p$-value of $0.45$, 
as compared to a $p$-value of $0.51$ for the pure SM (21~d.o.f.). 
 
\subsection*{\boldmath Extra loop contributions to the $gg$ and $\gam\gam$ couplings}

Let us now turn to case 2), where we allow for extra loop contributions to the $gg$ and $\gam\gam$ couplings of the Higgs boson, denoted $\dcg$ and $\dcp$, respectively. Such loops would involve particles beyond those present in the SM.  If these BSM particles receive mass from the Higgs mechanism, then they will give contributions to $\dcg$ and $\dcp$ that approach a constant value as their mass increases in size --- they do not decouple.  

Allowing for arbitrary $\dcg$ and $\dcp$ the minimum $\chi^2$ decreases from the SM fit value of $20.2$ to $12.0$ with the best values of $\dcg=-0.01$ and $\dcp=0.45$ for a non-zero value of the invisible rate, $\brinv=0.13$. The one-dimensional (1d) distribution of $\Delta\chi^2$ as a function of $\brinv$, after profiling over $\dcg$ and $\dcp$, is plotted as the dashed line in Fig.~\ref{fig:onedim}. The 95\% CL upper limit on the invisible rate increases dramatically to $\sim 60\%$.  
Larger $\brinv$ can be accommodated by increasingly large values of $\dcg$ so that the overall production rates in SM final states from $gg$ fusion processes remain the same. Rates in SM final states for  vector boson fusion (VBF) induced processes will decline somewhat and this is the primary reason for the increase in $\chi^2$ with increasing $\brinv$.

\begin{figure}[t!] \centering
\includegraphics[scale=0.46]{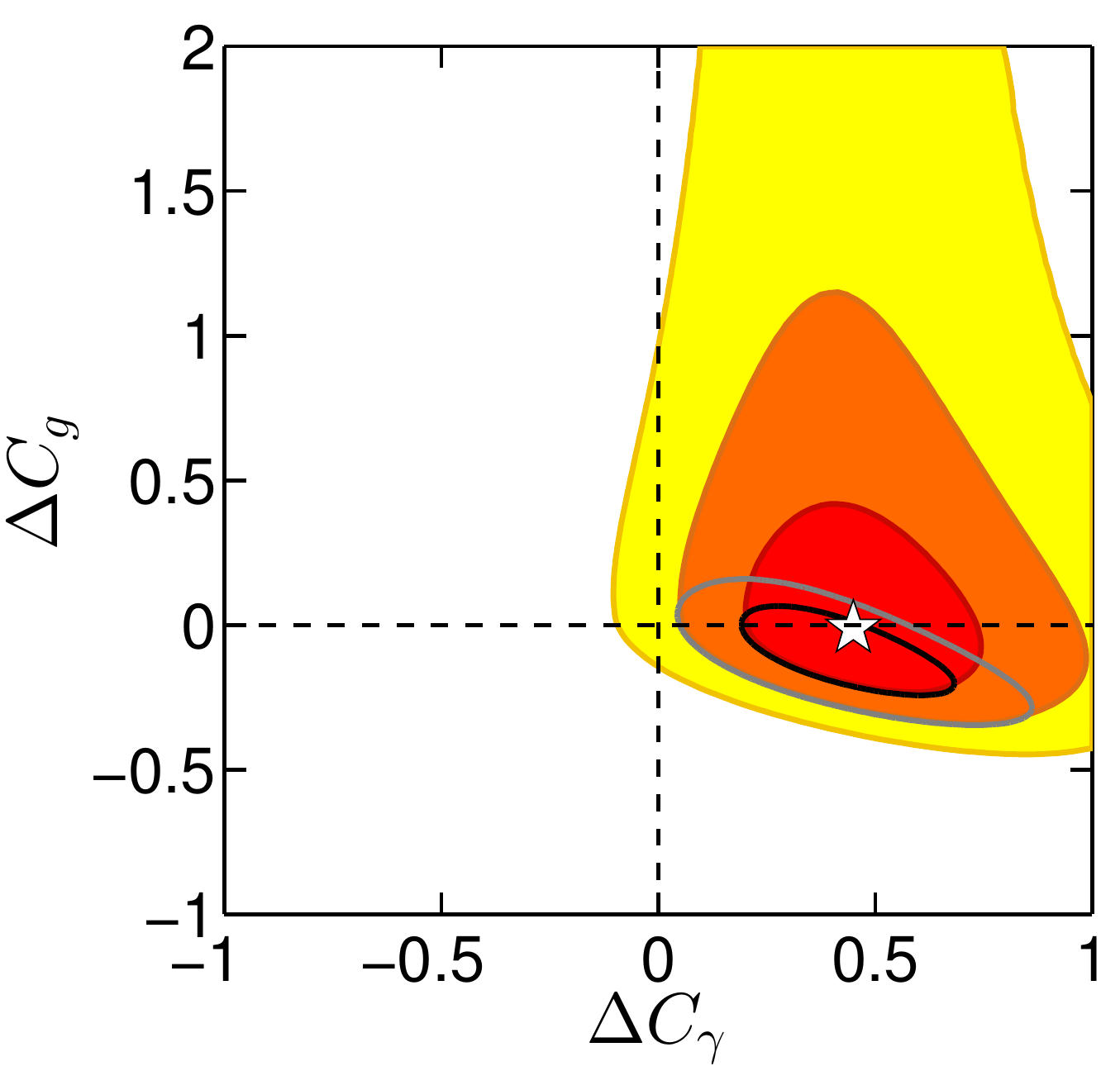}\\
\includegraphics[scale=0.46]{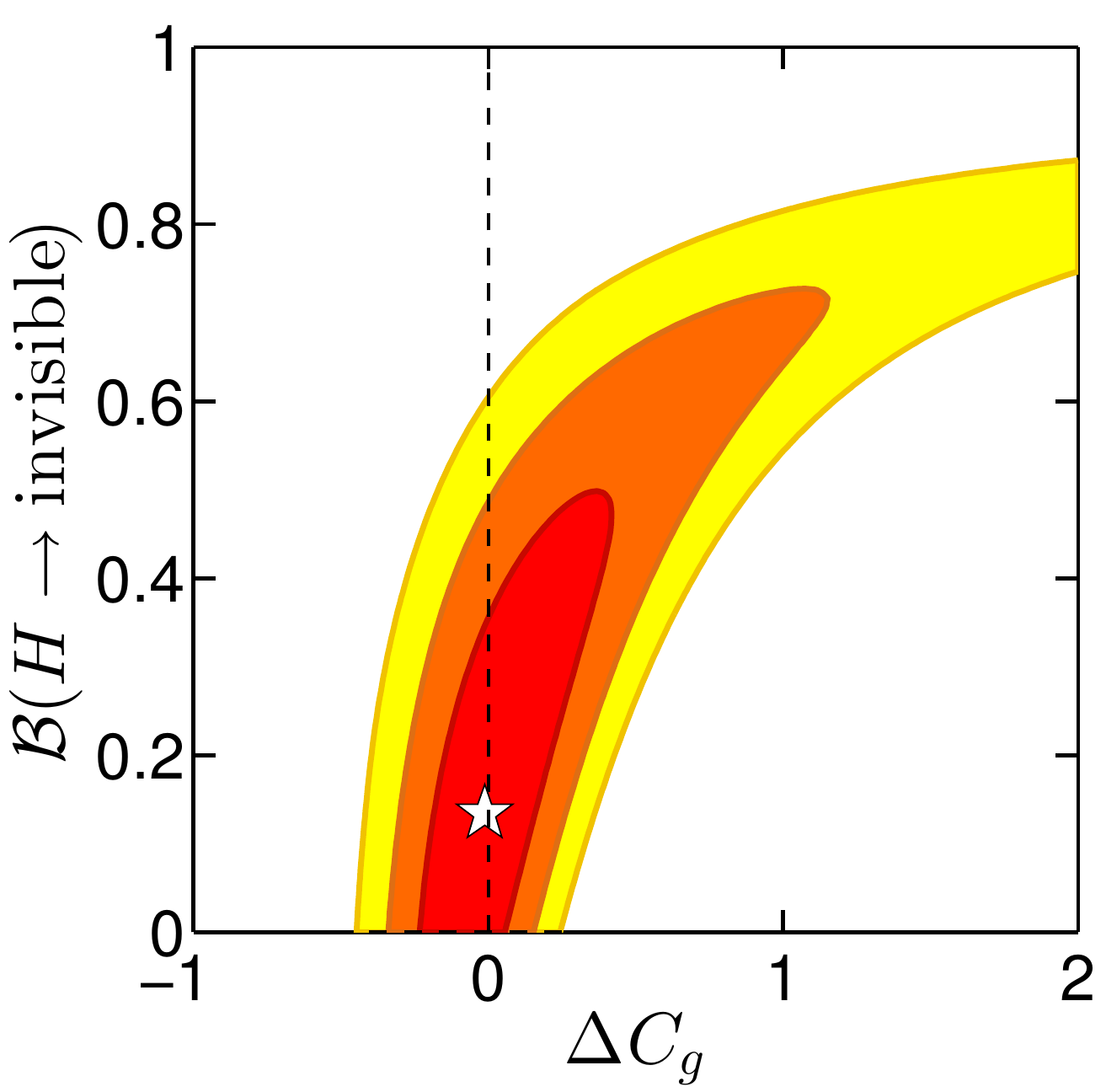}\quad
\includegraphics[scale=0.46]{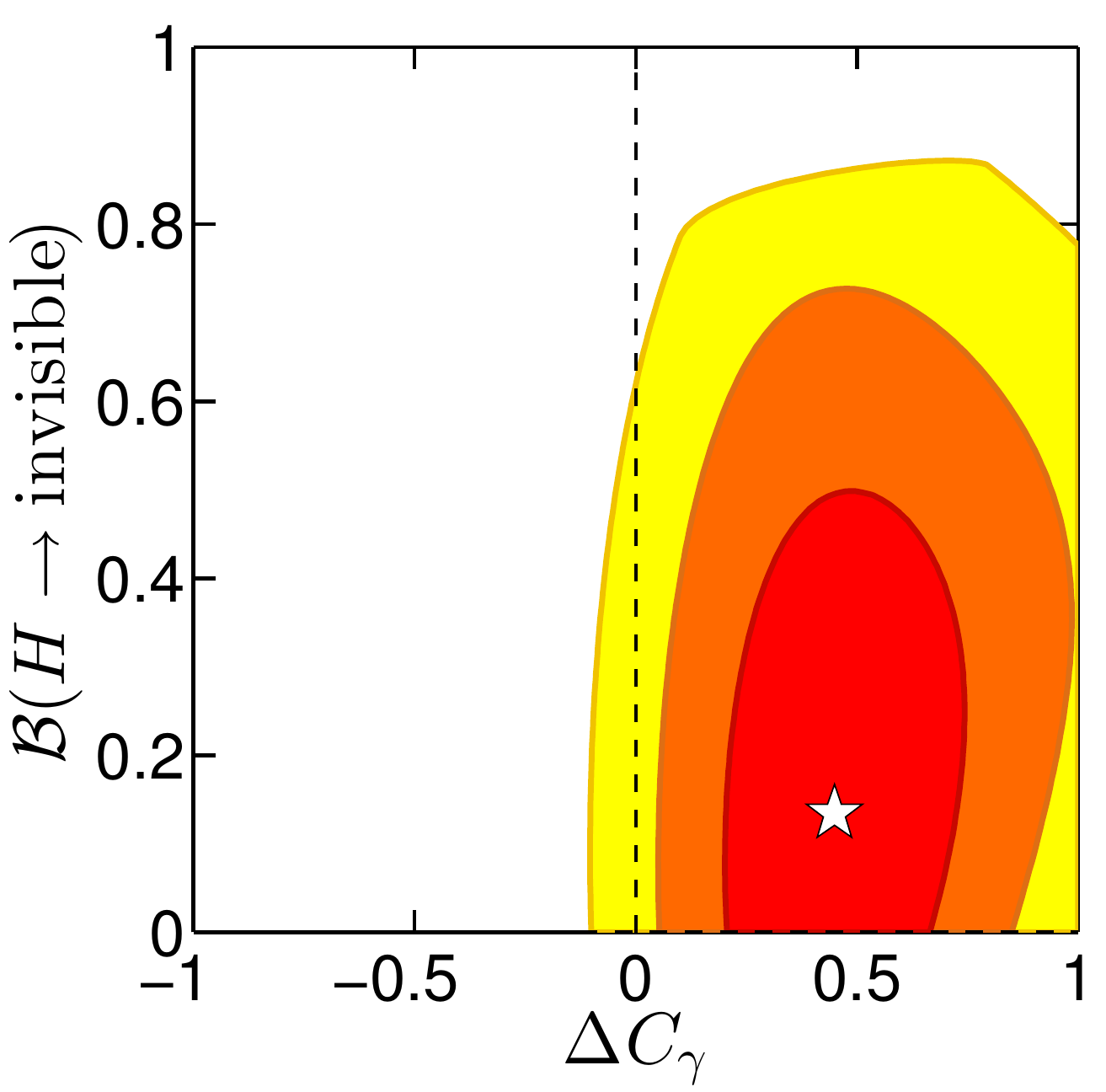}
\caption{$\brinv$ contours obtained by allowing for additional loop contributions $\Delta\cg$ and $\Delta\cp$ 
to the effective couplings of the Higgs to gluons and photons.
The red, orange and yellow areas 
(black, grey and light grey areas in greyscale print) are the 68\%, 95\% and 99.7\%~CL regions, respectively. 
The black and grey ellipses in the top plot show the 68\% and 95\% CL contours when invisible decays are absent. We find $\Delta\cg=-0.01_{-0.16}^{+0.24}$, $\Delta\cp= 0.45_{-0.17}^{+0.19}$ and 
$\brinv=0.13_{-0.13}^{+0.24}$ (where the errors correspond to  1d profiling). At the best fit point, marked as a white star, we find  $\chimin=12.03$ (for 18~d.o.f.). 
\label{fig:delta-cg-cp} }
\end{figure}

The $1\sigma$, $2\sigma$ and $3\sigma$ ranges in the  $\dcg$ versus $\dcp$ plane are shown in the top plot in Fig.~\ref{fig:delta-cg-cp}. 
It is interesting to compare to the case of arbitrary $\dcg$ and $\dcp$ without invisible decays, for which the $1\sigma$ and $2\sigma$ ranges are shown as black and gray contours in Fig.~\ref{fig:delta-cg-cp}. 
Allowing for invisible decays, we find 
$\dcg=-0.01_{-0.16}^{+0.24}$, $\dcp=0.45_{-0.17}^{+0.19}$ ($p\simeq 0.85$) as compared to 
$\dcg=-0.09\pm{0.10}$, $\dcp=0.43_{-0.16}^{+0.17}$ ($p\simeq 0.87$) when invisible decays are absent. 
Contours of $\brinv$ versus $\dcg$ and versus $\dcp$ are shown in the bottom row of  Fig.~\ref{fig:delta-cg-cp}. 
We see that, as mentioned, larger $\brinv$ can be accommodated by increasing $\dcg$. On the other hand, 
larger negative $\dcg$, as for example induced by light stops with large mixing, disfavors invisible decays.
Finally, $\brinv$ is rather insensitive to $\dcp$. 

\subsection*{\boldmath Couplings to fermions and/or gauge bosons deviating from 1}

In case 3), we allow for arbitrary $\cu,\cd,\cv$ while varying $\brinv$.  We do not allow for extra contributions from BSM particles to the $gg$ and $\gam\gam$ Higgs couplings, \ie\ we assume $\cg=\anti\cg$ and $\cp=\anti\cp$.  
It is interesting to consider several sub-cases.

\begin{figure}[t] \centering
\includegraphics[scale=0.46]{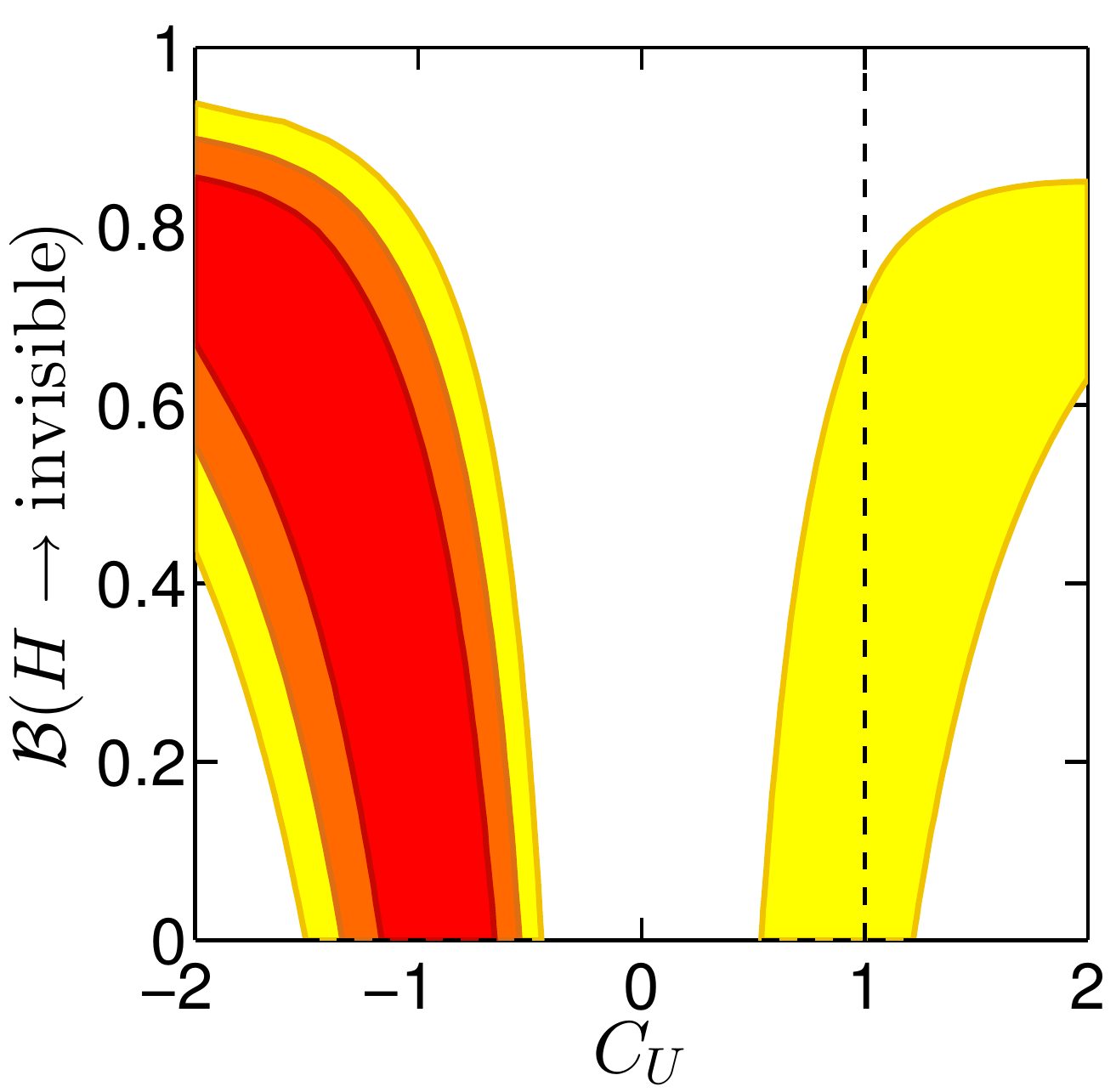}\quad
\includegraphics[scale=0.46]{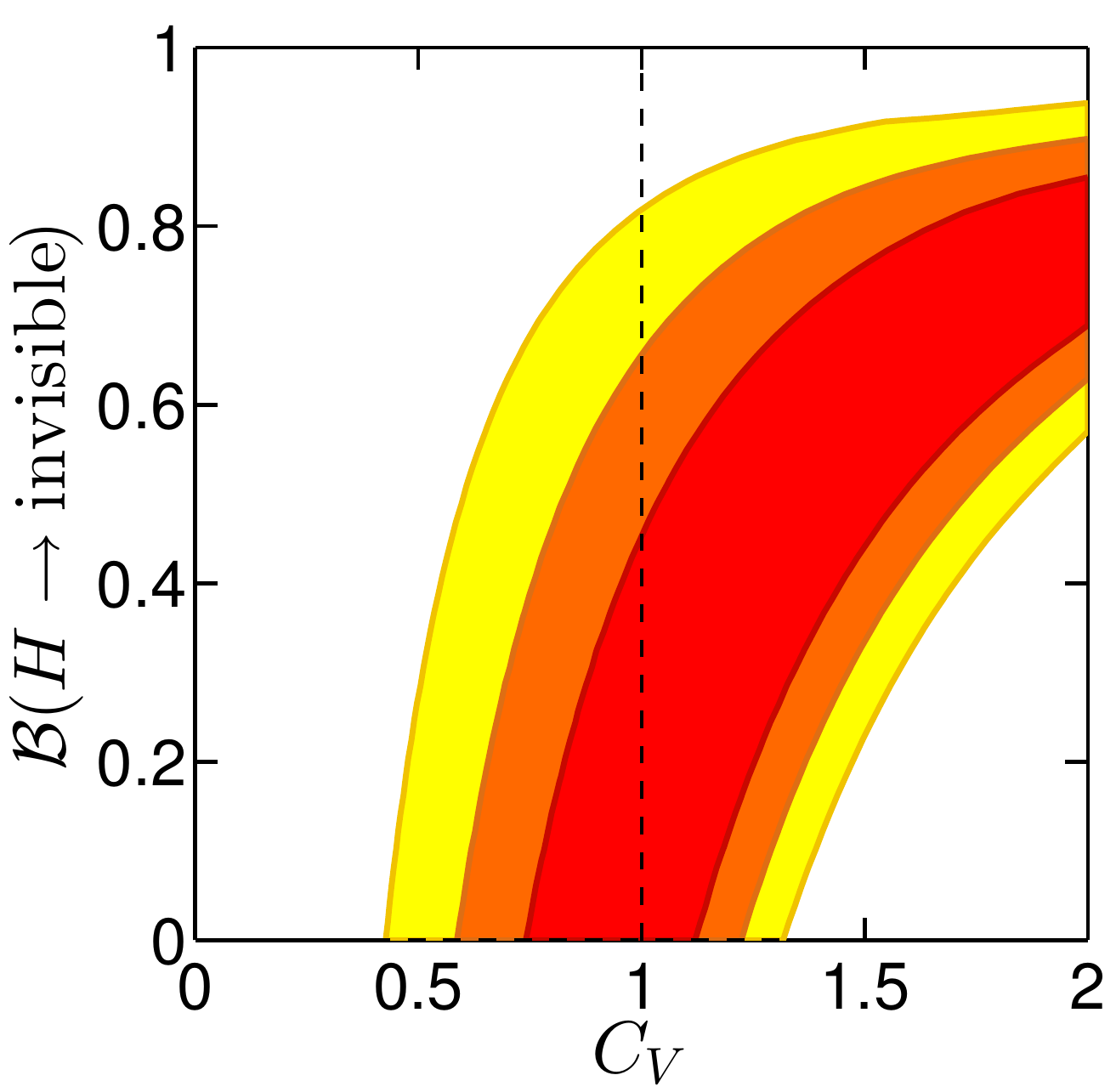}
\caption{Fit of $\brinv$ allowing for deviations of $\cu$, $\cd$, $\cv$ from 1,    
but without extra loop contributions, \ie\ $\dcg=\dcp=0$. 
Same color code as in Fig.~\ref{fig:delta-cg-cp}. 
Allowing up to 100\% deviations in $\cu$ and $\cv$, the $2\sigma$ ($3\sigma$) limit is $\brinv<0.88$ (0.92).
\label{fig:cu-cd-cv} }
\end{figure}

For general $\cu,\cd,\cv$ (limited however to the range $\pm2$), we find $\cu=-0.86_{-1.14}^{+0.14}$, 
$\cd=0.99_{-0.26}^{+1.01}$ and $\cv=0.95_{-0.13}^{+1.05}$. 
The best fit point has $\chimin=11.95$ for 17 d.o.f.\ and is basically
the same as without invisible decays.  A large $\brinv$ can yield a
small $\Delta \chi^2$ when simultaneously increasing both $|\cu|$ (so as
to increase the $gg$ production rates) and $\cv$ (so as to increase the
VBF production rates) in order to compensate for the decreasing
branching ratios to SM particle final states. (We also need to increase
$|C_D|$ in order to have enough decays into $b\bar{b}$ and $\tau\tau$.)
As in the fits of~\cite{Belanger:2012gc},  the minimum $\chi^2$ is
achieved for negative $\cu$, something that is very difficult in the
context of most theoretical models, see the comment at the end of this Section.  
In addition, the large values of $\cv>1$ required for a good fit with
large $\brinv$ imply 
an enhancement of the isospin $I = 2$ cross section, which may be
achieved if the Higgs state at $\sim$ 125 GeV mixes
with higher SU(2) representations~\cite{Logan:2010en,Falkowski:2012vh}.
(Within extensions of the
 Higgs sector by an arbitrary number of SU(2) doublets and/or singlets
 only, one generally obtains $C_V\leq 1$.)
 Allowing for large
negative $\cu$ and large positive $\cv$  (up to $\cu\sim-2$, $\cv\sim
2$), $\brinv$ can only be constrained to $< 0.88$ at 95\% CL,   see
Fig.~\ref{fig:cu-cd-cv}.

\begin{figure}[t] \centering
\includegraphics[scale=0.46]{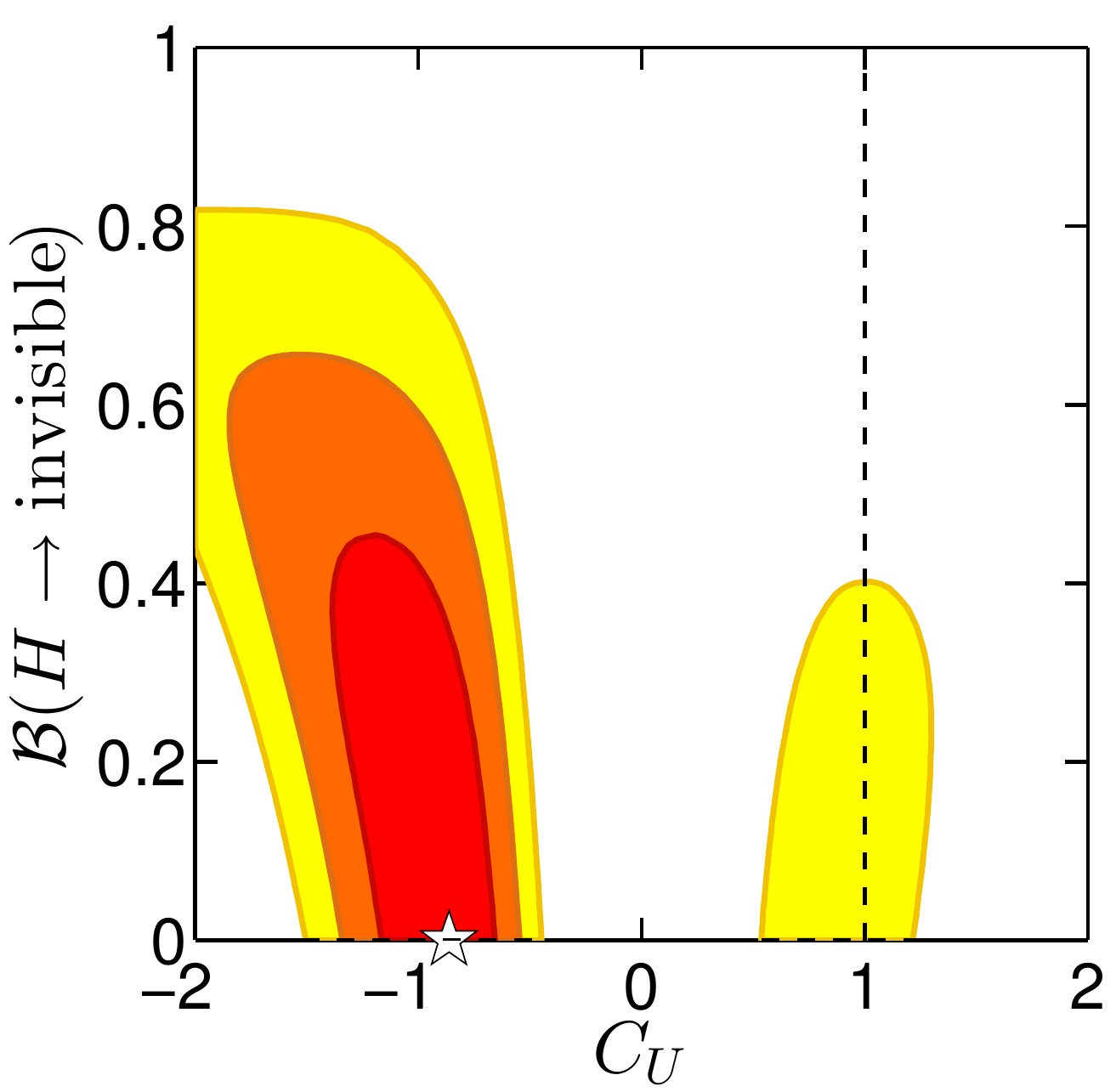}\quad
\includegraphics[scale=0.46]{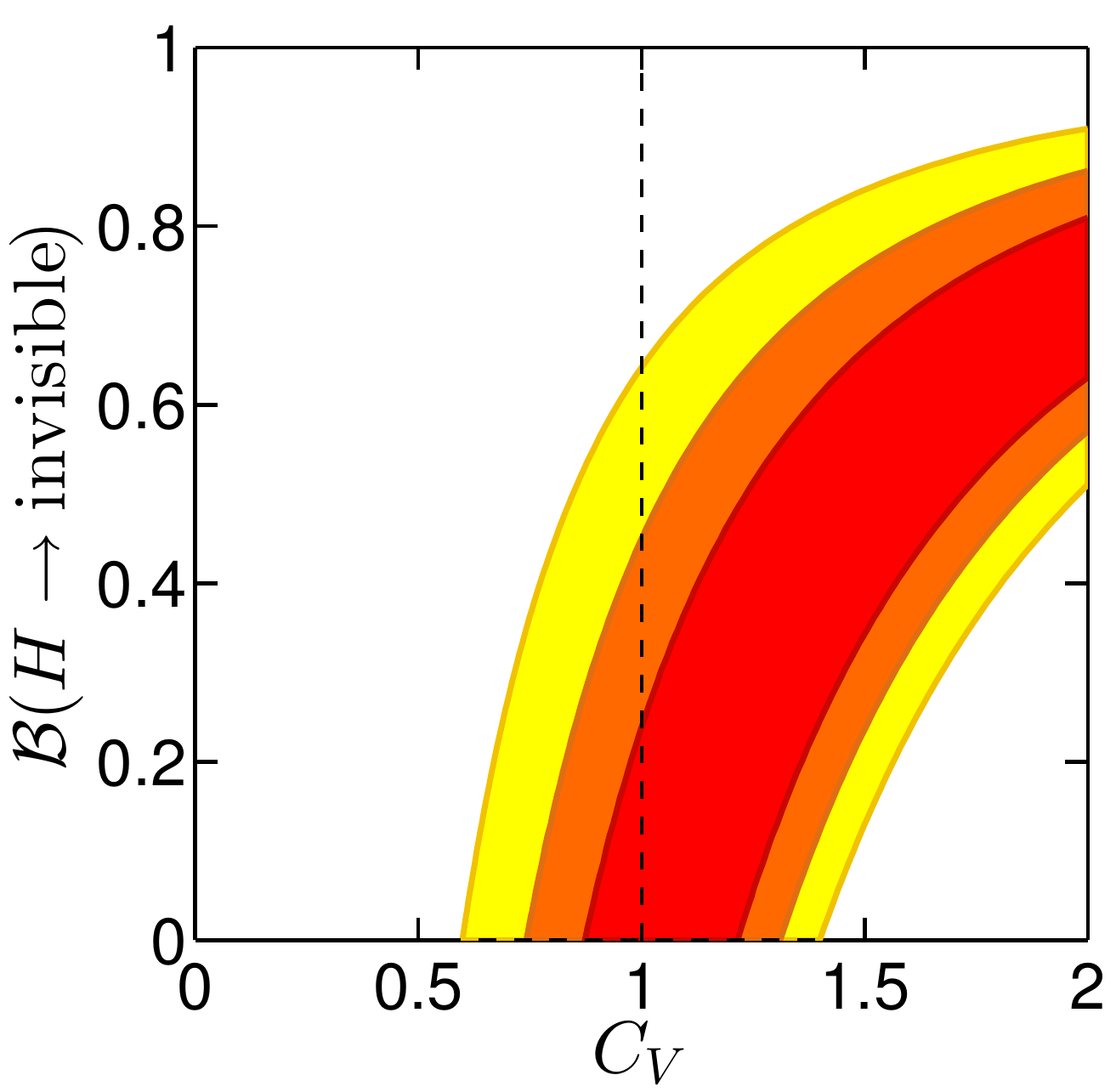}
\caption{Same as Fig.~\ref{fig:cu-cd-cv} but restricting either $\cv\le1$ (left) or $\cu>0$ (right). 
\label{fig:cu-cd-cv-2} }
\end{figure}

Restricting the fit to $\cv\le1$, as relevant for a Higgs sector consisting of doublets+singlets,  
the minimum $\chi^2$ remains at $11.95$ but the $\Delta\chi^2$ increases rapidly with the invisible rate. 
We find $\cu = -0.86_{-0.34}^{+0.14}$ in this case, and $\brinv<0.56$ at 95\%~CL. 
Obviously, imposing $\cv\leq 1$ not only greatly restricts the production cross sections that can be achieved in vector boson fusion but also restricts the two-photon partial width of the Higgs and therefore decreases the value of $\brinv$ that can allow reasonable consistency with the experimental observations. 
$\brinv$ versus $\cu$ for $\cv\le1$ is shown in the left plot in Fig.~\ref{fig:cu-cd-cv-2}.

Another interesting case is to require $\cu>0$ while allowing $\cv>1$. 
In this case, the minimum $\chi^2$ point is characterized by $\cu=0.85_{-0.13}^{+1.11}$, $\cd=0.85_{-0.21}^{+1.11}$, $\cv=1.05_{-0.12}^{+0.95}$, and 
the fit worsens to $\chimin=18.7$. Further, $\brinv$ can only be constrained to $< 0.84$ at 95\%~CL. 
For illustration, see the right-hand-side plot in Fig.~\ref{fig:cu-cd-cv-2}. 

\begin{figure}[t] \centering
\includegraphics[scale=0.46]{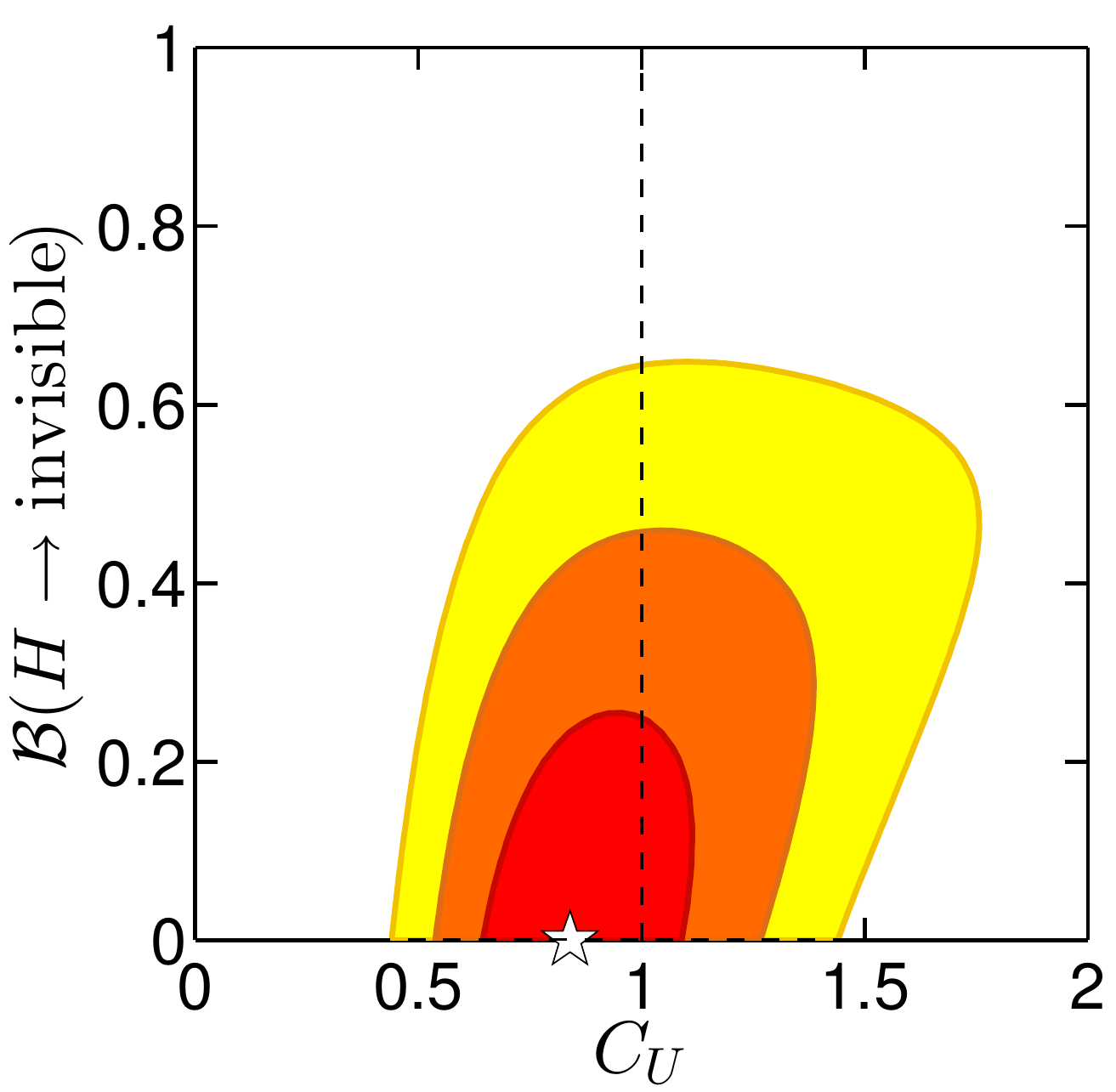}
\includegraphics[scale=0.46]{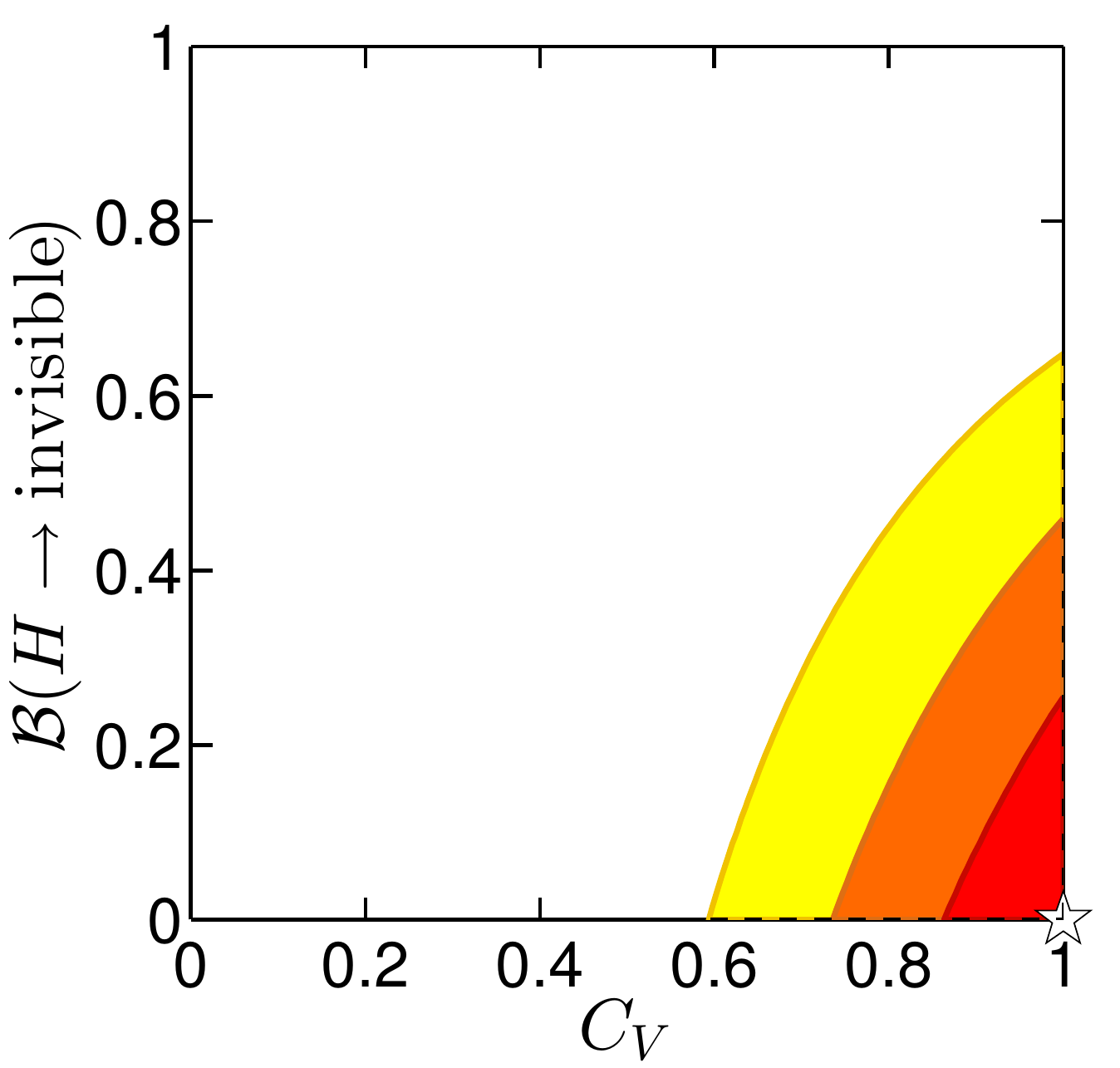}
\caption{$\brinv$ contours obtained by requiring $\cv\le1$ and $\cu,\cd>0$. 
Same color code as in previous figures.
\label{fig:brinv-cu-cd-pos-cvle1} }
\end{figure}

Perhaps most interesting from the theoretical point of view is the case in which $\cu,\cd>0$ and $\cv\le1$ is required.  
With these constraints, $\chimin=18.9$ (\ie\ not far from the SM fit value of $20.2$) and the 95\% CL limit on $\brinv$ is $\sim 0.36$. The $\Delta\chi^2$ distribution for this case is shown as dash-dotted line in Fig.~\ref{fig:onedim}. 
For $\brinv$ versus $\cu$ and $\cv$, see Fig.~\ref{fig:brinv-cu-cd-pos-cvle1}.
 
\begin{figure}[t] \centering
\includegraphics[scale=0.4]{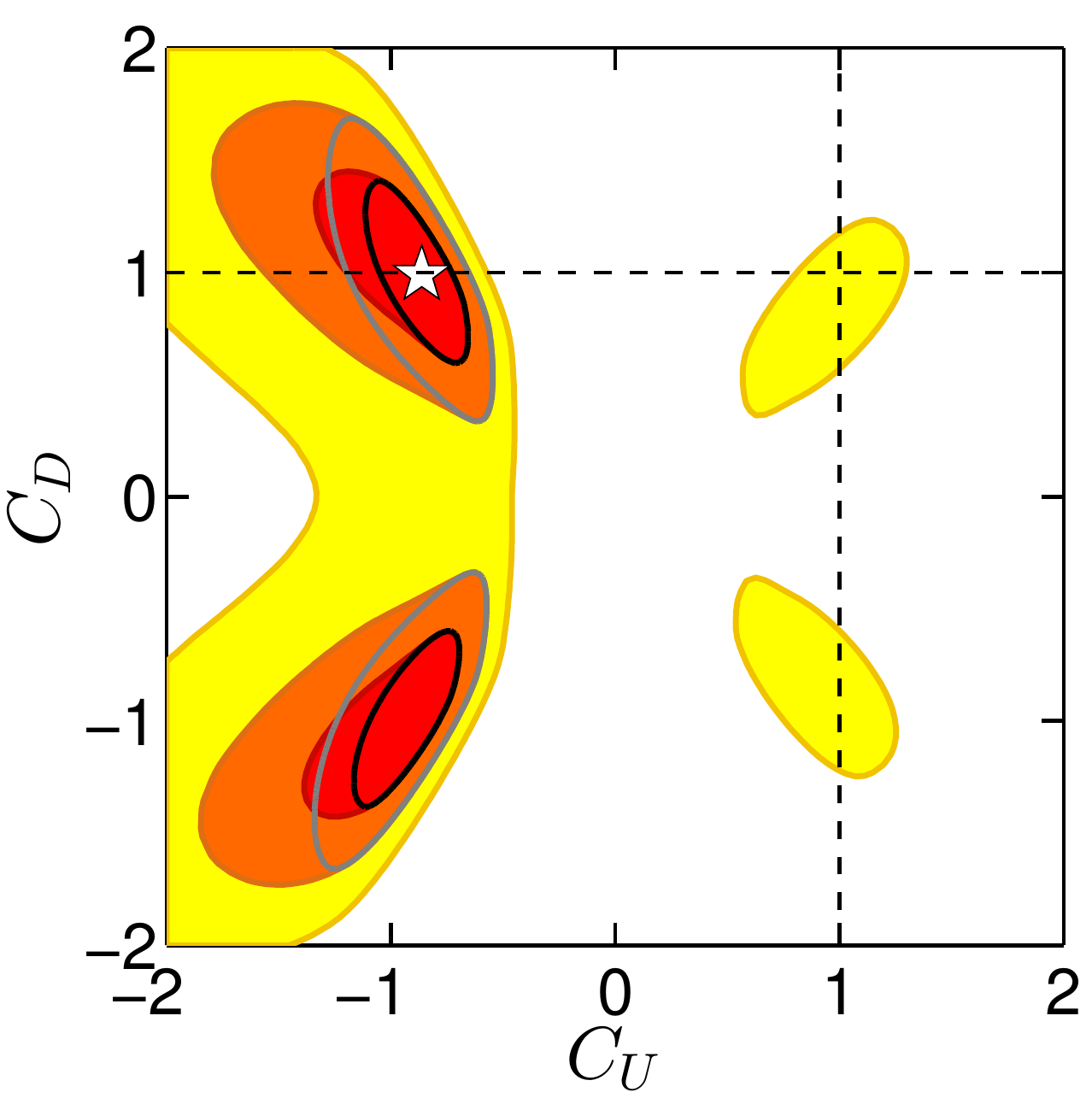}
\includegraphics[scale=0.4]{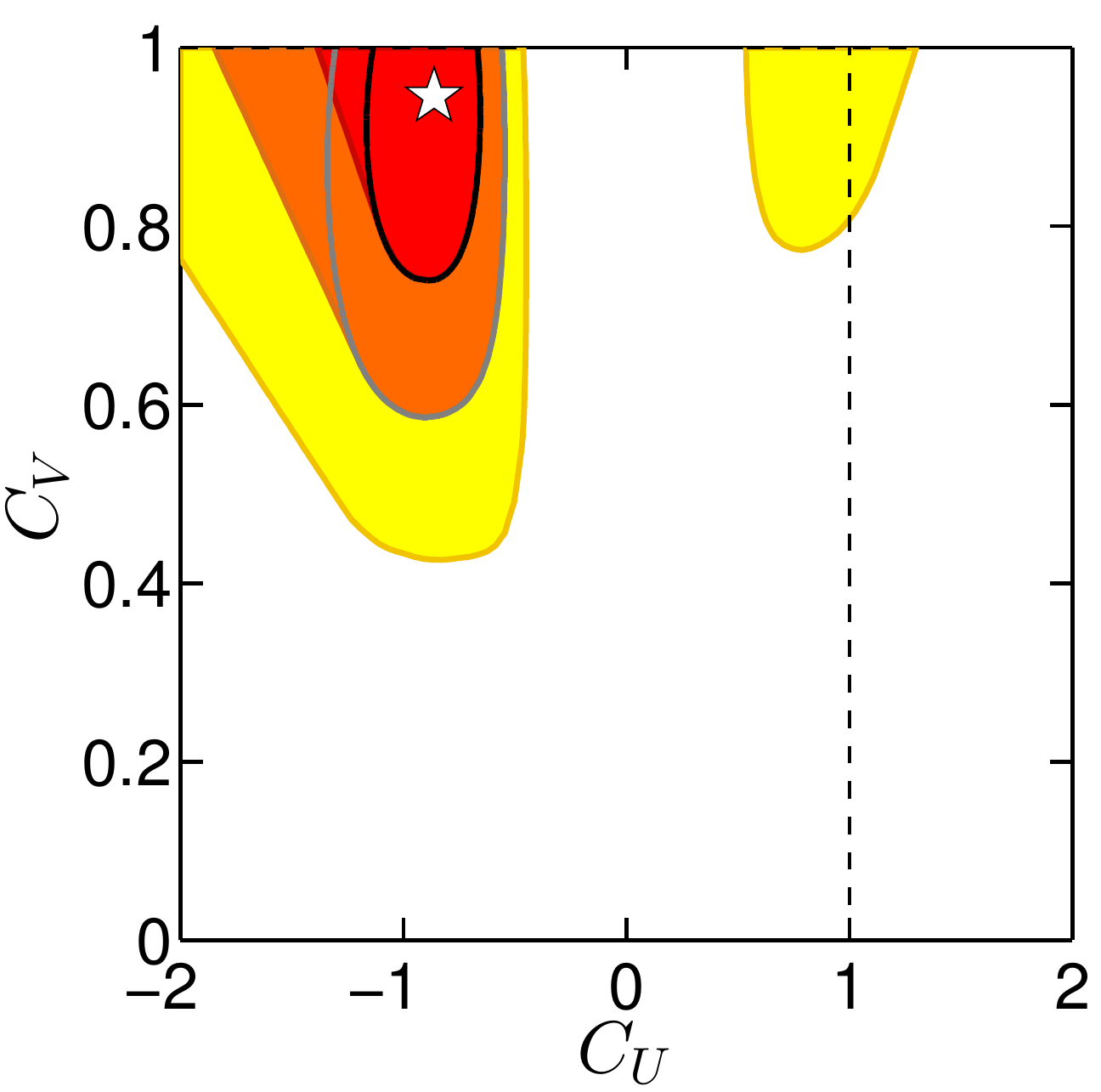}
\includegraphics[scale=0.4]{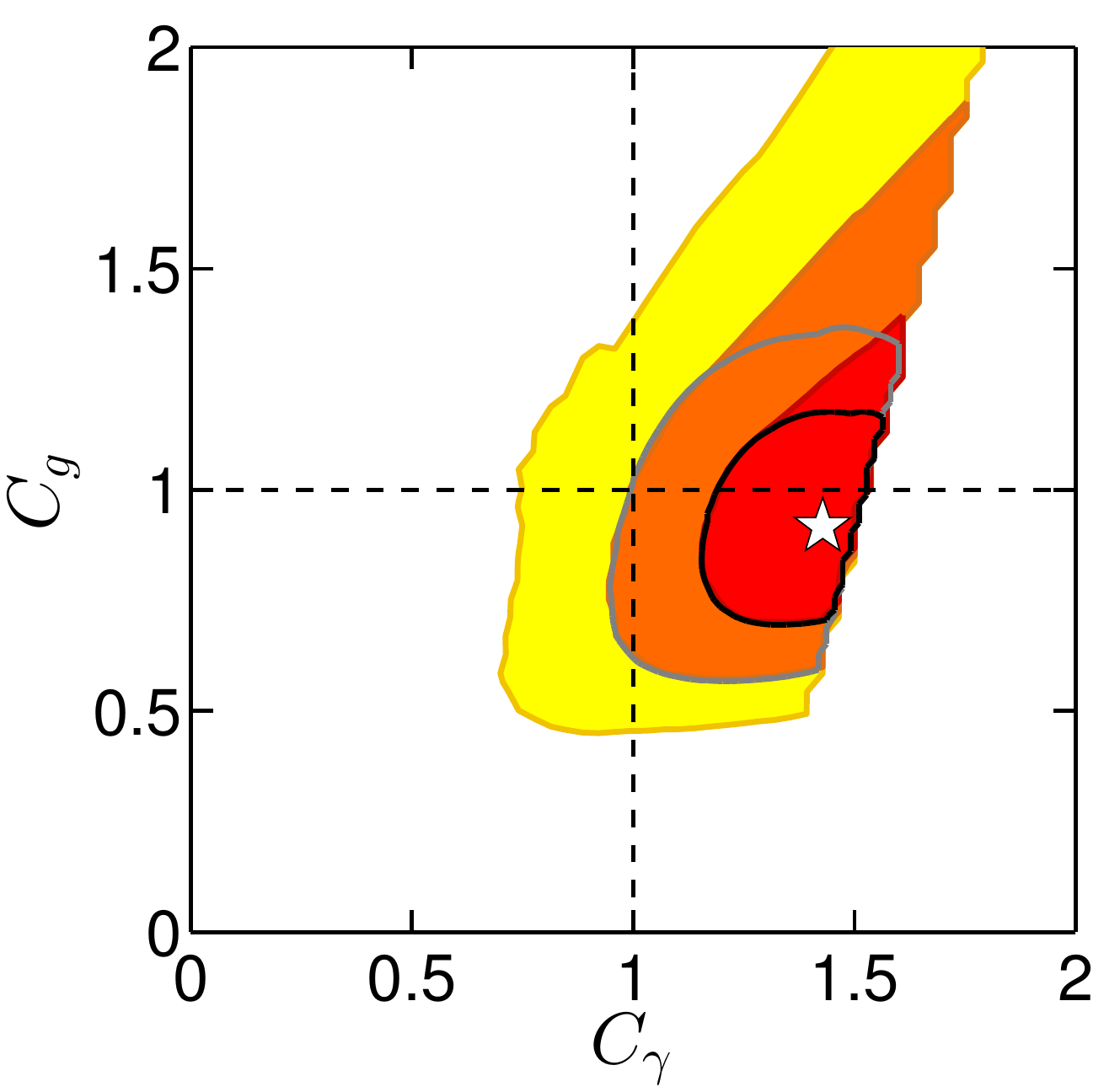}\\[2mm]
\includegraphics[scale=0.4]{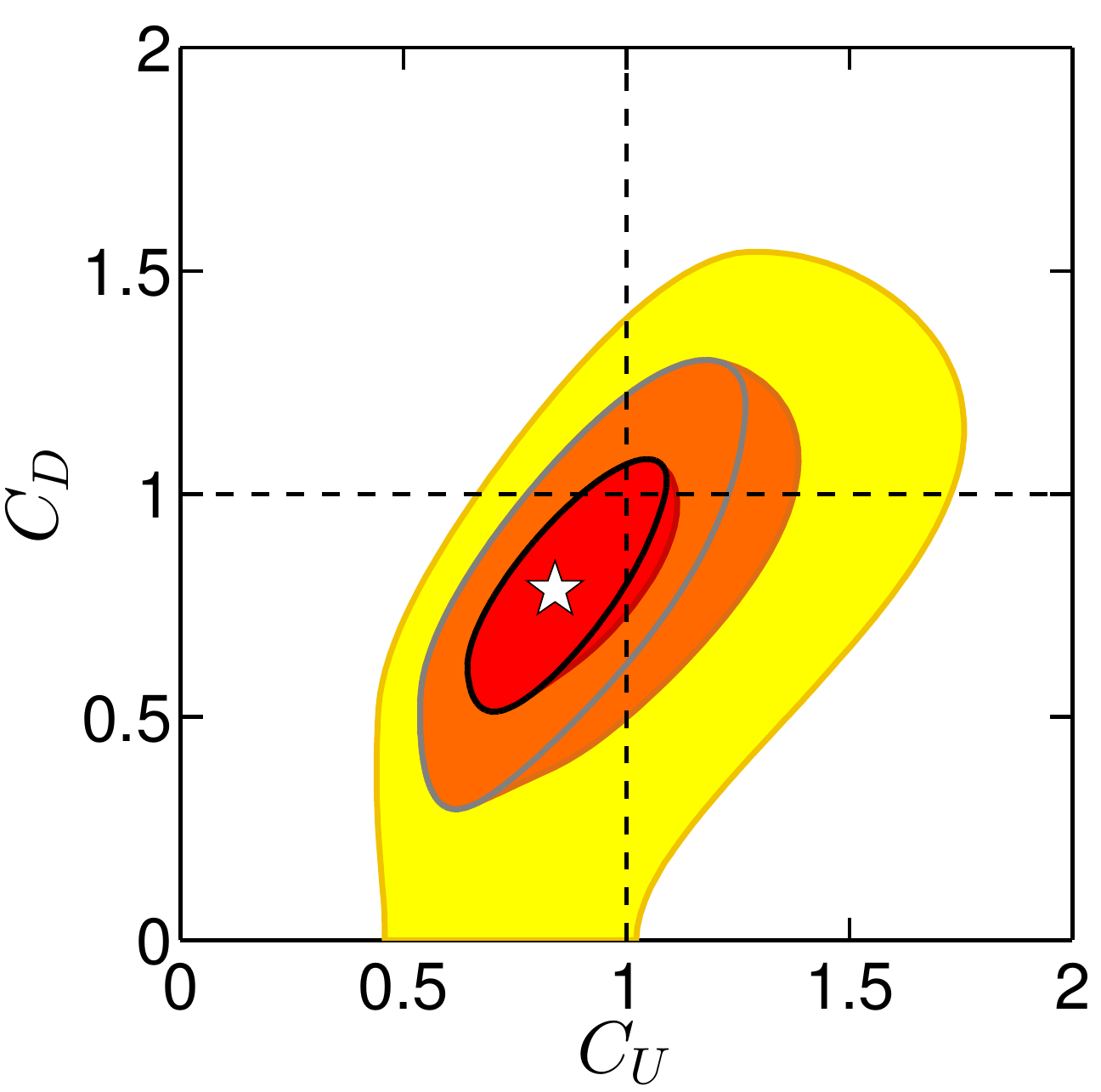}
\includegraphics[scale=0.4]{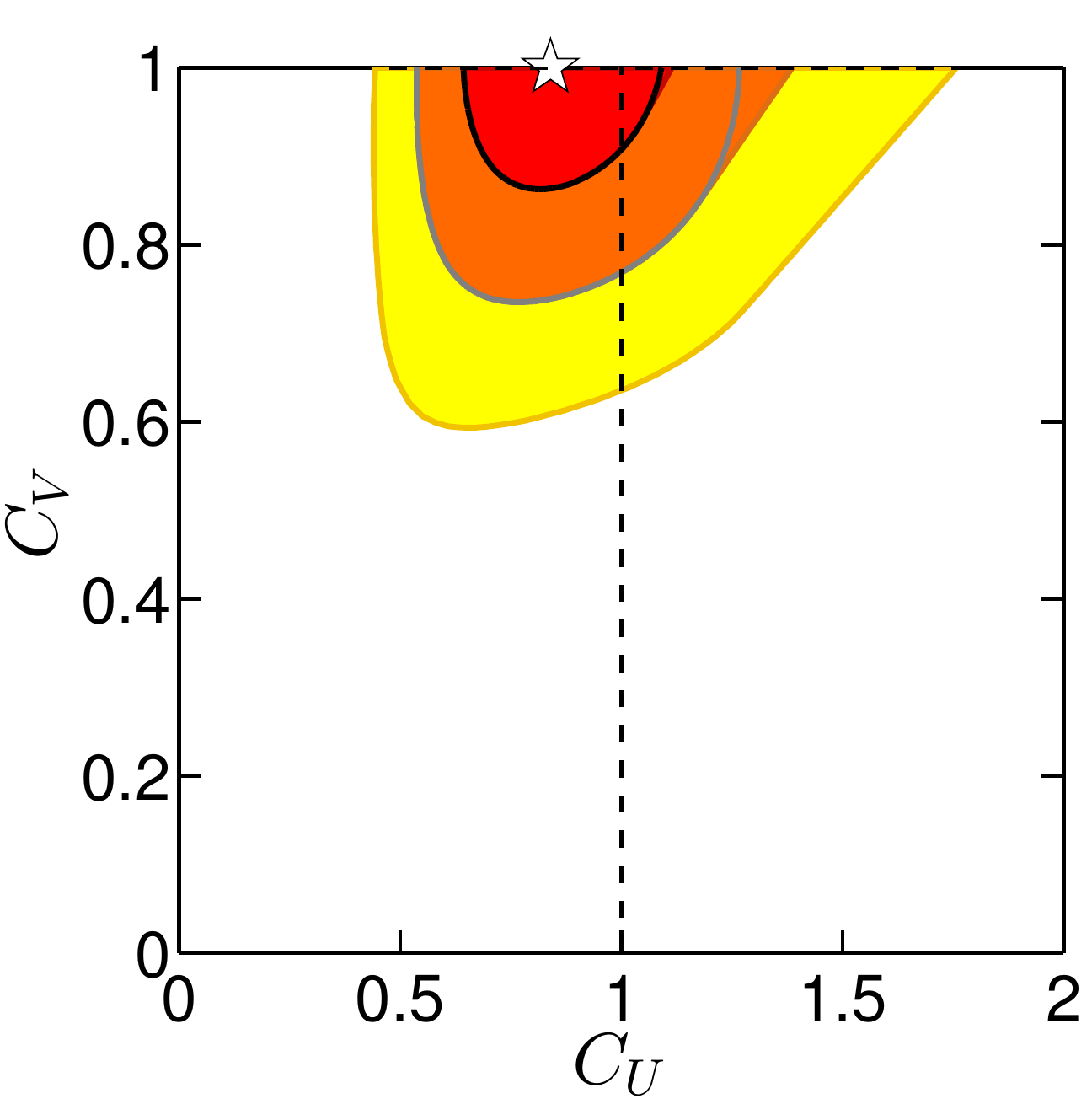}
\includegraphics[scale=0.4]{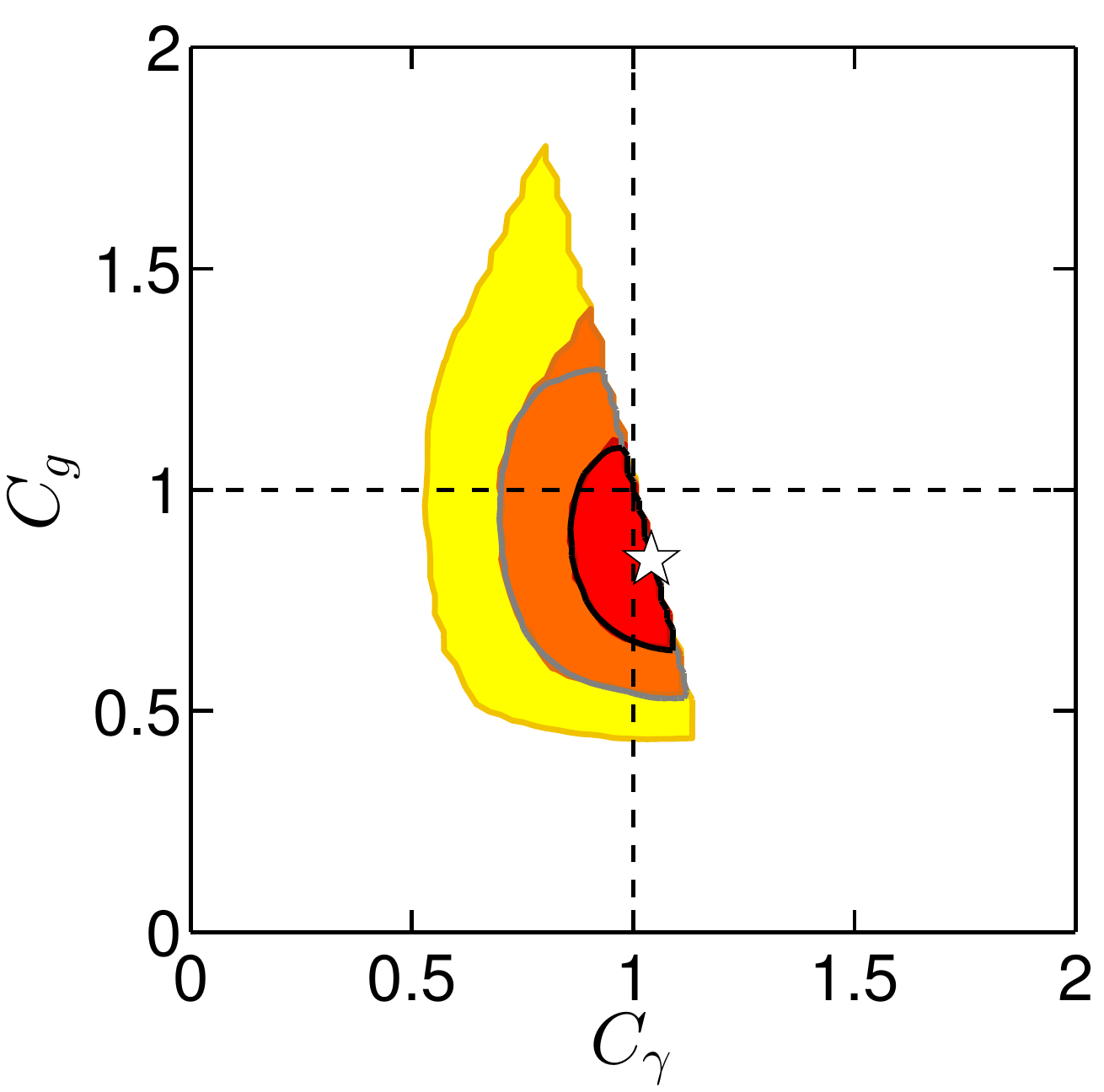}
\caption{Fit of $\brinv$ allowing for deviations of $\cu$, $\cd$, $\cv$ from 1.    
In the top row of plots, $\cv\le 1$ but $\cu,\cd$ may have either sign. 
In the bottom row, $\cv\le1$ and $\cu,\cd>0$. 
Color code {\it etc.}\ as in previous figures.
\label{fig:cu-cd-cv-par} }
\end{figure}

The relevant 2d correlations between parameters, illustrating the discussion above, are shown in Fig.~\ref{fig:cu-cd-cv-par} for:  the fit requiring $\cv\le 1$ but allowing arbitrary signs for $\cu,\cd$ (top row); and the fit requiring both $\cv\le1$ and $\cu,\cd>0$ (bottom row). 
In order to see the impact of invisible decays on the coupling fits, we have superimposed the $1\sigma$ and $2\sigma$ regions  from \cite{Belanger:2012gc} obtained for $\brinv=0$. 

Let us end this section with a comment on $C_U < 0$.
A negative sign of $C_U$---while maintaining a positive sign of
$m_t$---is actually not easy to achieve. If the top quark and Higgs
bosons are considered as fundamental fields, it would require that the
top quark mass is induced dominantly by the vev of at least one
additional Higgs boson which is not the Higgs boson considered here, and
leads typically to various consistency problems as discussed, \eg, in
\cite{Choudhury:2012tk}.

\section{Further probes of invisible or undetected Higgs decays}

Truly invisible Higgs decays can be probed at the LHC in monojet searches
in either the ggF mode where a gluon is radiated from the initial state,
or in VBF when one of the jets is missed. 
Invisible decays can also be probed in $ZH$ associated production with $Z\to l^+l^-$. 
In~\cite{Djouadi:2012zc}, sensitivity to the monojet searches is phrased in terms of limits on 
\begin{equation}
   \rinv(X)={\sigma(X\to H)\brinv\over \sigma_{\rm SM}(X\to H)}\,. 
\end{equation}
A 95\% CL upper limit of $\overline{\rinv}={2\over 3}\, \rinv({\rm ggF})+{1\over 3}\,\rinv({\rm VBF})<1.3$ was obtained using the CMS monojet analysis at $\sqrt{s}=7$~TeV and ${\cal L}=4.7$~fb$^{-1}$~\cite{Chatrchyan:2012me}.  The relative contributions of the gluon and vector boson fusion production mechanisms were assumed to be the same as in the SM after the analysis cuts.  Of course, this need not apply if $\cu,\cd,\cv$ are allowed to vary.
(Assuming only one production channel, the 95\% CL upper limits are $\rinv({\rm ggF})<1.9$ and $\rinv({\rm VBF})<4.3$.) 
The projected limit for $\sqrt{s}=8$~TeV and ${\cal L}=15$~fb$^{-1}$ is $\overline{\rinv}<0.9$.
\begin{figure}[t!] \centering
\includegraphics[scale=0.45]{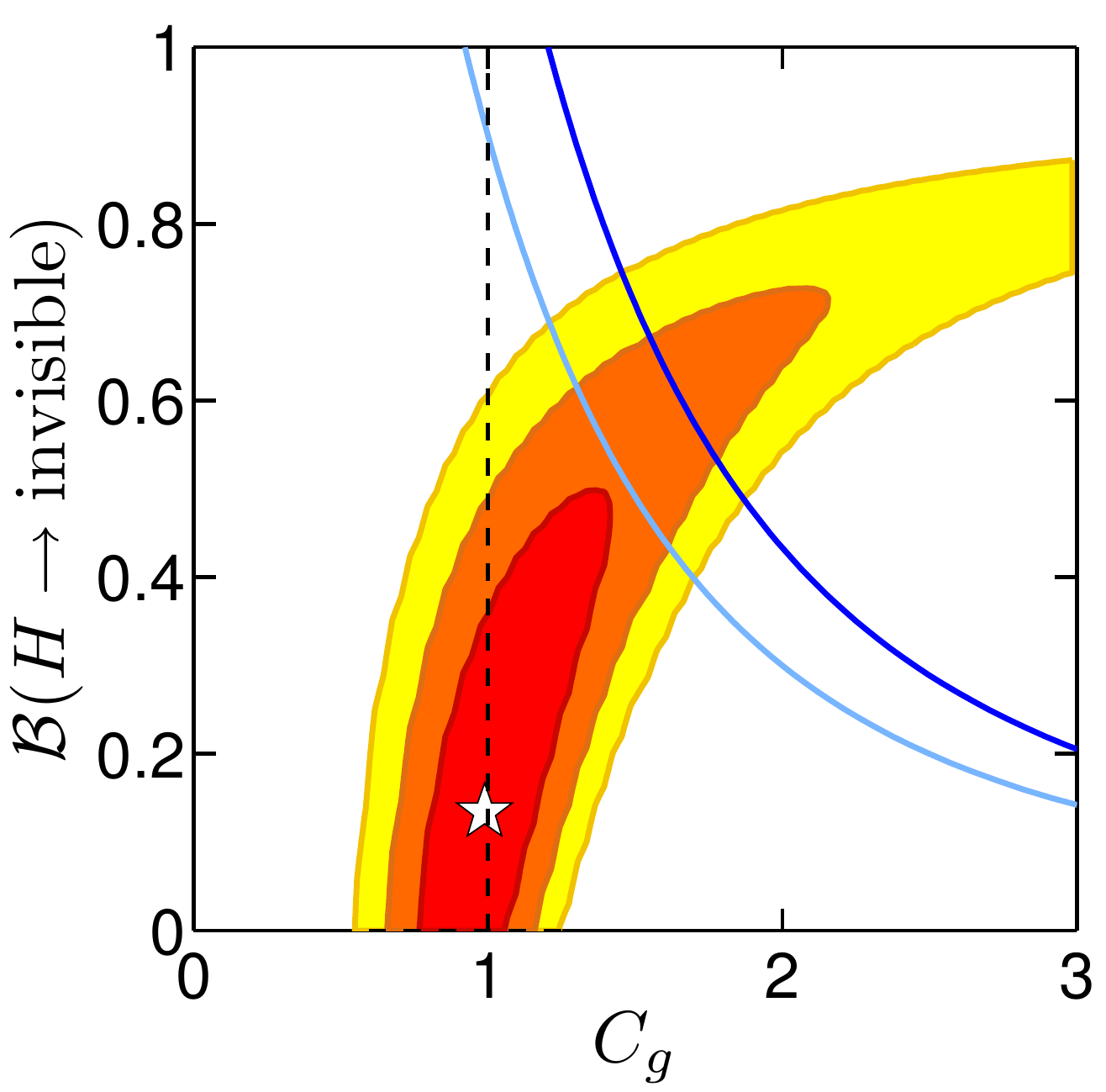}\quad
\includegraphics[scale=0.45]{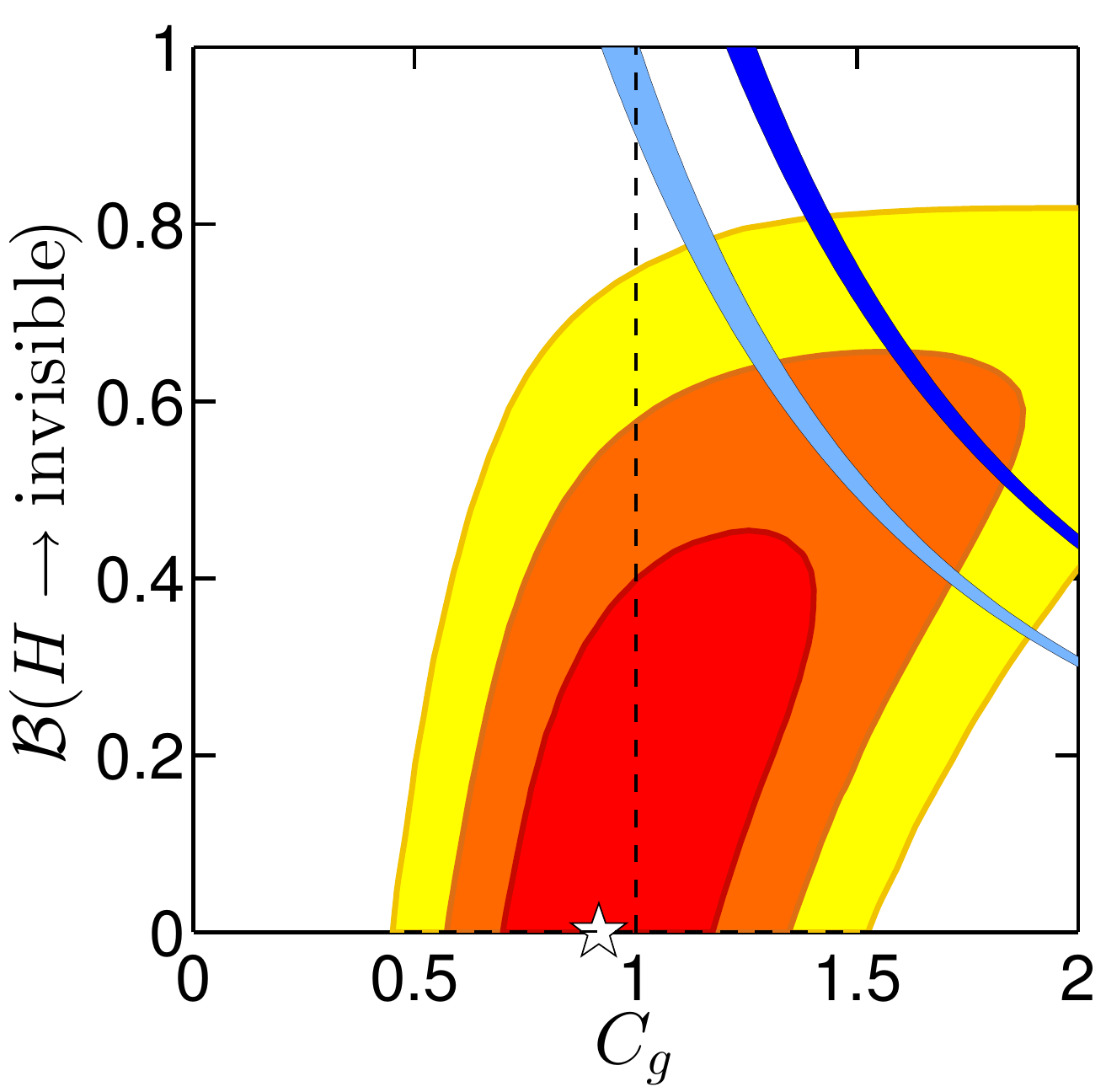}\quad
\caption{Contours of $\brinv$ versus $\cg$, on the left for the fit of
$\dcg$ and $\dcp$,  on the right for the fit of $\cu,\cd,\cv$ with
$\cv\le 1$.  The red, orange and yellow regions are 68\%, 95\% and $99.7$\%
CL regions, respectively.  The dark (light) blue bands show the
constraints from the monojet search, $\overline \rinv<1.3$ ($0.9$) at 95\% CL,
with the bands  obtained by varying $\dcp$ (left plot) or $\cv$ (right
plot) within their fitted $2\sigma$ ranges. 
\label{fig:monojet} }
\end{figure}
Since the signals are also
proportional to the possibly non-standard Higgs production cross
section $\sim C_g^2$ in the ggF mode or $\sim C_V^2$ in the VBF mode, 
$\rinv({\rm ggF})= C_g^2\, \brinv$ and 
$\rinv({\rm VBF})= C_V^2\,\brinv$. 
Upper limits on $\rinv({\rm ggF})$ will thus
constrain $\brinv$ as function of $\cg$, as shown in 
Fig.~\ref{fig:monojet} for case 2) on the left and case 3) with
$\cv\le1$ on the right.  These plots should be compared to $\brinv$
versus $\dcg$ in Fig.~\ref{fig:delta-cg-cp} and  $\brinv$ versus $\cu$
in Fig.~\ref{fig:cu-cd-cv-2}.
The dark (light) blue bands indicate
$\overline \rinv({\rm ggF})<1.3$ ($0.9$), with the band obtained by  varying
$\dcp$ or $\cv$ within $2\sigma$.  As can be seen, the monojet searches
are already quite complementary in constraining  invisible Higgs decays 
when there is a large increase in the production cross section. 

Another analysis~\cite{Ghosh:2012ep} considered searching for invisible
Higgs in the 2 jets and missing $p_T^{}$ channel showing that a  $5\sigma$ signal could be
observed at 8~TeV for ${\cal L}=20$~fb$^{-1}$ for an SM production cross section provided $\brinv>0.84$, 
while the LHC at 14~TeV  could  probe $\brinv>0.25$  with ${\cal L}=300$~fb$^{-1}$. 
Since the 2 jets $+p_T^{\rm miss}$ channel is dominated by VBF production (86\% after analysis cuts), this can be useful 
to constrain the cases with $C_V >1$. For example, 
$\brinv> 0.23$ $(0.4)$ could be probed at 8~TeV for $C_g=1$ and $C_V=2$ $(1.5)$, thus covering a large fraction of the currently allowed parameter space at large values of $C_V$ in Fig.~3 and Fig.~4 (right panel).

Let us finally comment on decays that may in principle be detectable. 
The ability of the LHC to probe for Higgs decays into light pseudoscalars, $H\to AA$, depends on the decays of the $A$'s. 
The most likely $A$ decays are $A\to b\anti b$,  dominant for $m_A>2m_b$, and $A\to \tau^+\tau^-$, dominant for $2m_\tau<m_A<2m_b$.  A review, with detailed referencing, of the possibilities for the LHC in various production modes in the cases of these decays is given in \cite{Almarashi:2011te}.  In two-Higgs-doublet models $A\to q\anti q, gg,\ldots$ (where $q$ is a light quark, \eg\ $s$ or $c$) can be significant if $\tanb\lsim 1.7$ or $m_A<2m_\tau$. LHC sensitivity in this case has been examined for $ZH$ production in~\cite{Englert:2012wf}.   In all the different $A$ decay scenarios pretty much full LHC luminosity, ${\cal L}=100-1000\fbi$ at $\rts=14\tev$ is required to place strong limits (\eg\ $\br<10\%$ at 95\% CL) on $H\to AA$ decays.

Another potentially interesting decay channel, that may have escaped observation, is $H\to\gamma+E_T^{\rm miss}$, 
with a soft photon. This may arise for instance in $H\to \tilde\chi^0_1\tilde G$ decays followed by 
$\tilde\chi^0_1\to\gamma\tilde G$, where $\tilde G$ denotes a gravitino \cite{Petersson:2012dp} or a 
goldstino \cite{Dudas:2012fa}.

\section{Interplay with direct dark matter searches}

Assuming that the invisible particle which the Higgs potentially decays into is the dark matter of the Universe, 
the LHC bounds on $\brinv$ can be turned into bounds on the DM scattering off nucleons, mediated by Higgs exchange, {\it cf.}~\cite{Burgess:2000yq,Kanemura:2010sh,He:2011de,Mambrini:2011ik,Fox:2011pm,Djouadi:2011aa}. These bounds are often much stronger than the current limits from XENON100 for $m_\chi<62$~GeV ({\it i.e.}\ $m_H/2$). 
Both the invisible width of the Higgs and the spin-independent cross-section for scattering on protons depend on the 
square of the Higgs--DM--DM coupling $C_{\rm DM}$.  
If the DM is a Majorana fermion, $\chi$, the invisible width arising from $H\to \chi\chi$ decays is given by
\begin{equation}
   \Gamma_{\rm inv}=\Gamma(H \rightarrow \chi\chi) =\frac{g^2}{16\pi} m_H C_\chi^2 \beta^3 \, ,
\end{equation}
where $\beta=(1-4m_\chi^2/m_H^2)^{1/2}$ and $C_\chi$ is defined by ${\cal L}=gC_\chi \bar{\chi}\chi H$. 
In case of the DM being a real scalar, $\phi$, we have ${\cal L}=g m_\phi C_\phi \phi\phi H$ and 
\begin{equation}
   \Gamma_{\rm inv}=\Gamma(H \rightarrow \phi\phi) =\frac{g^2}{32\pi} \frac{m_\phi^2C_\phi^2}{m_H} \beta \, .
\end{equation}

The spin-independent cross-section for scattering on a nucleon, considering only the Higgs exchange diagram, 
can then be  directly related to the invisible width of the Higgs: 
\begin{equation}
   \sigma_{\rm SI}= \eta \mu_r^2 m_p^2  \frac{g^2}{M_W^2}\, \Gamma_{\rm inv} 
   \left[  C_U (f_u^N+f_c^N+f_t^N) +C_D (f_d^N+f_s^N+f_b^N)+ \frac{\Delta C_g}{\what C_g}f_g^N\right]^2 \, 
   \label{eq:sigSI}
\end{equation}
with $\eta=4/(m_H^5 \beta^3)$ for a Majorana fermion and 
$\eta=2/(m_H^3 m_\phi^2\beta)$ for a real scalar;  
$\mu_r$ is the reduced mass and $f_q^N\,(f_g^N)$ are the quark (gluon) coefficients in the nucleon. 
We take the values  $f_s^p=0.0447$, $f_u^p=0.0135$, and $f_d^p=0.0203$ 
from an average of recent lattice results~\cite{Junnarkar:2013ac,Belanger:2013oya}. The gluon and heavy quark ($Q=c,b,t$) coefficients are related to those of light quarks, and $f_g^p= f_Q^p= 2/27(1-\sum_{q=u,d,s} f_q^p)$ at leading order. 
Since the contribution of heavy quarks to the scattering amplitude originates from their contribution to the $Hgg$  coupling, 
we write the effect of $\dcg$, the last term in eq.~(\ref{eq:sigSI}), in terms of an additional top quark contributing to the $Hgg$ coupling; numerically  $\what C_g=\overline C_g=1.052$ with only the SM top-quark contribution taken into account for computing $\overline C_g$.

For the numerical evaluation of $\sigma_{\rm SI}$, we use micrOMEGAs~\cite{Belanger:2008sj,Belanger:2013oya} in which the relation between the heavy quark coefficients and the light ones are modified by QCD corrections.  This amounts to taking
\begin{equation}
C_Q f_Q^p\rightarrow  C_Q \left (1+\frac{35 \alpha_s(m_Q)}{36 \pi} \right) f_Q^p \,, \qquad \Delta C_g f_g^p \rightarrow \Delta C_g \left(1-\frac{16 \alpha_s(m_t)}{9 \pi} \right) f_g^p \,.
\end{equation}

\begin{figure}[t]\centering
\includegraphics[scale=0.5]{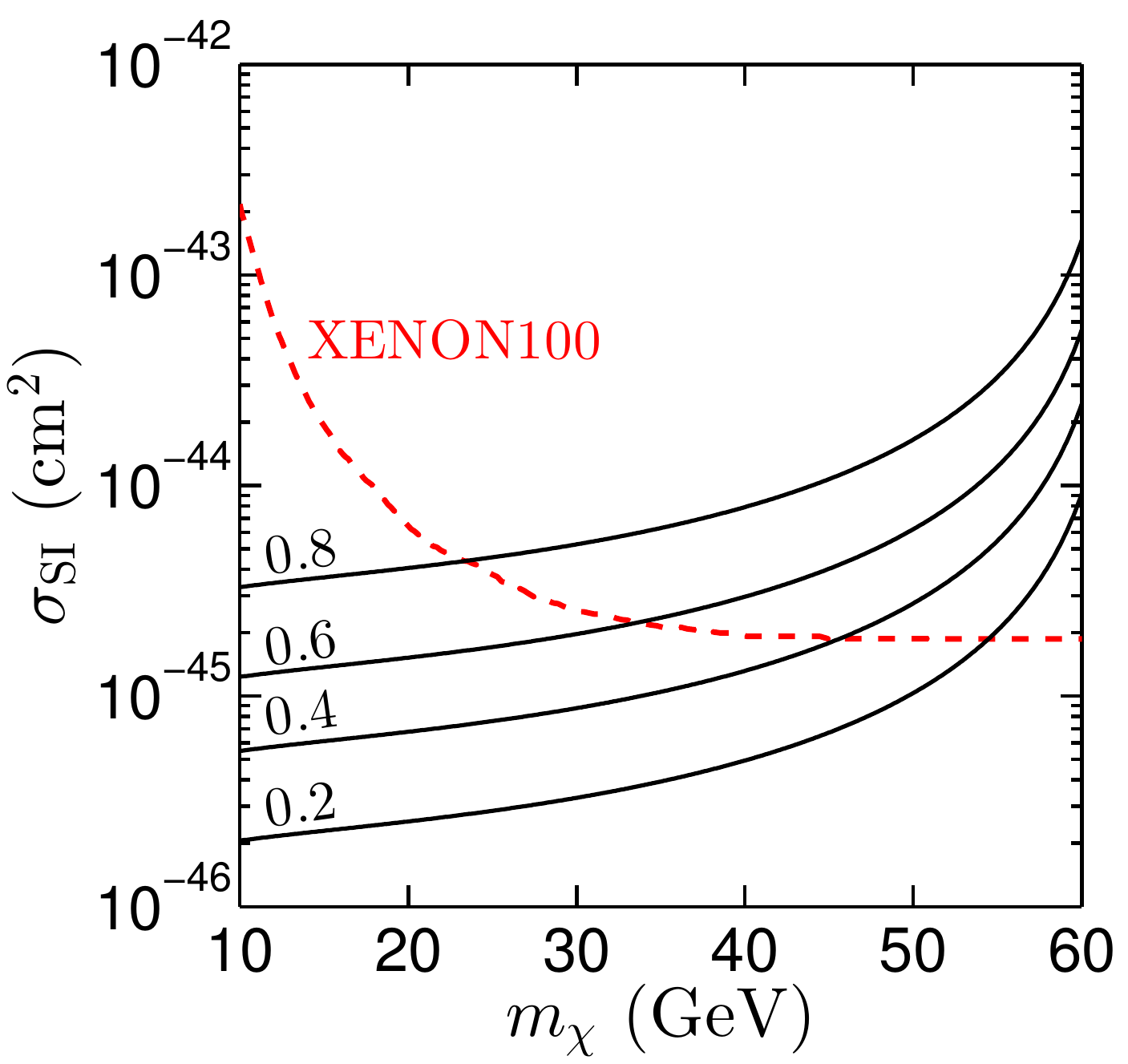}\quad\includegraphics[scale=0.5]{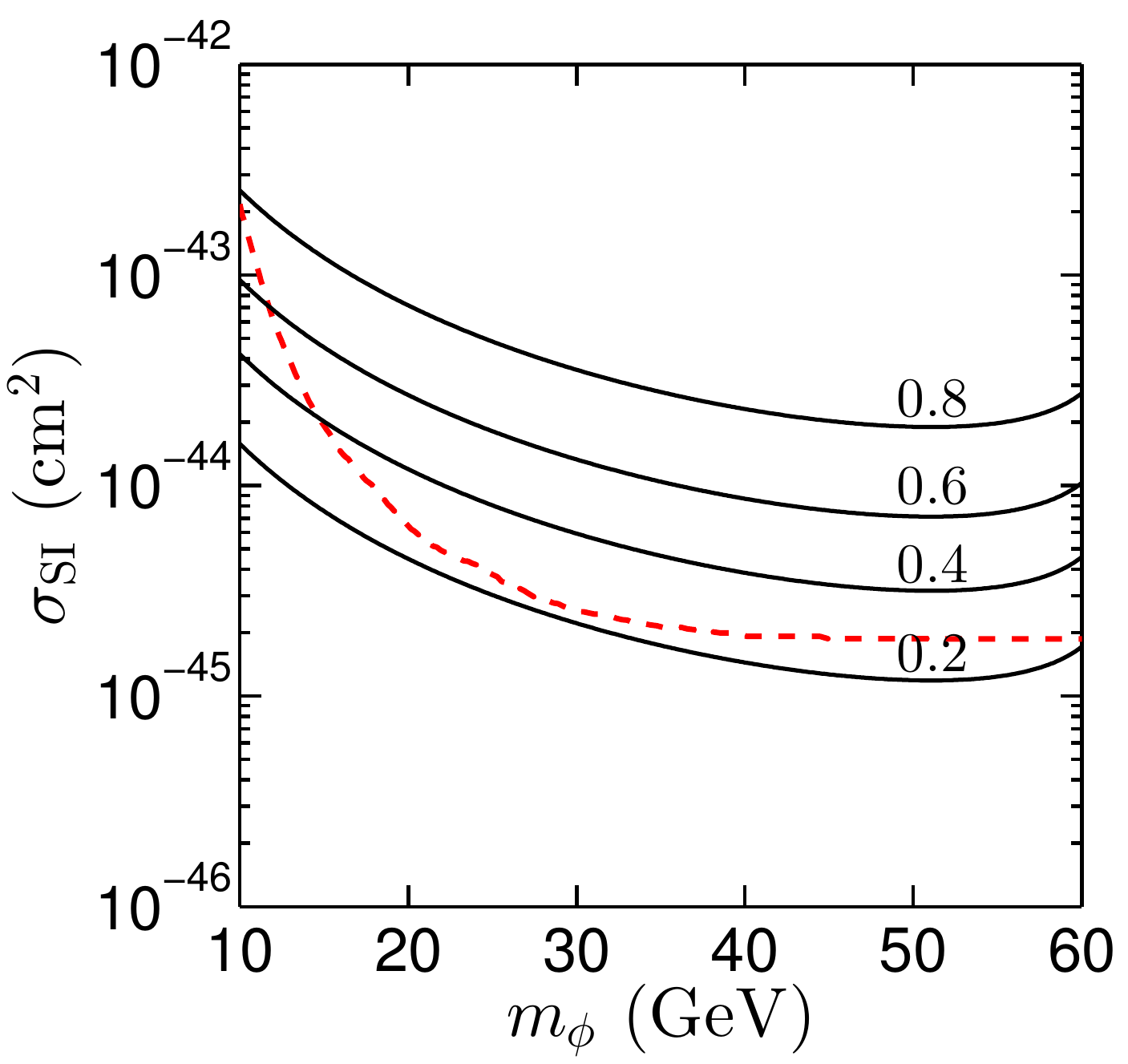}
\caption{$\sigma_{\rm SI}$ as a function of the mass of the DM particle, for  $\brinv=0.2,0.4,0.6,0.8$ (from bottom to top) for the case of a Majorana $\chi$ (left panel) or a real scalar $\phi$ (right panel) when $C_U=C_D=C_V=1$ and $\Delta\cg=\Delta\cp=0$, \ie\ an SM Higgs plus invisible decays. 
The red dashed curves show the XENON100 exclusion limit.
\label{BRinv-sigmaSI} }
\end{figure}

The results for $\sigma_{\rm SI}$ versus the DM mass and for different $\brinv$ are displayed in Fig.~\ref{BRinv-sigmaSI} for a Majorana fermion (left panel)\footnote{For a Dirac fermion, the cross sections are a factor 1/2 smaller.}
and a real scalar (right panel) assuming $C_U=C_D=C_V=1$. 
As can be seen,  for a Majorana fermion the current XENON100 limits~\cite{Aprile:2012nq} exclude, for example, 
$\brinv>0.4$ when $46~{\rm GeV} < m_\chi < m_H/2$. For scalar DM, the cross sections are larger, and 
XENON100 excludes $\brinv>0.4$ for $m_\phi \gtrsim 15$~GeV. 
These limits become much stronger when $C_U$ and/or $C_D$ are large provided they have the same sign. 
Further, these limits become stronger  when we include a non-zero value of $\Delta C_g$. For example, for  $\Delta C_g=1$  we find that  $\sigma_{\rm SI}$ increases by a factor 1.8 as compared to the case $\Delta C_g=0$  for any given value of $\brinv$. This increase is due in part to the new contribution in eq.~(\ref{eq:sigSI}) and in part because  a larger coupling of the DM to the Higgs is necessary to maintain a constant value of $\brinv$.
Note that  imposing universality of quark couplings to the Higgs  has an impact on our  predictions for 
$\sigma_{\rm SI}$ since all quark flavors contribute to this observable,  whereas  universality plays basically no role for Higgs decays as only the third generation is important.  
 
When $C_U<0$ and $C_D>0$, corresponding to the best fit for case 3),   there is a destructive interference between the $u$-type and $d$-type quark  contributions such that $\sigma_{\rm SI}$ is much below the current limit. Note however that a negative sign of $C_U$  would require that $m_t$ is induced dominantly by the vev of a Higgs boson
which is not the Higgs boson considered here; if such a Higgs boson also couples to dark matter it could then contribute significantly to the SI cross section. Without a complete  model for the Higgs sector it is therefore difficult to make generic predictions in this case.
  
When the DM candidate is a Dirac fermion and one assumes the same amount of matter and anti-matter in the early universe, the results for $\sigma_{\rm SI}$ are  simply a factor $1/2$ lower then those obtained in  the Majorana case. However if this fermion also couples to the $Z$, this gives an additional positive contribution to  $\sigma_{\rm SI}$, thus leading to stronger constraints from direct detection experiments.  Similar arguments hold for the case of a complex scalar, as compared to a real scalar.

\section{Conclusions}

Assuming that the 125~GeV state observed at the LHC is, indeed, a Higgs boson a very important question is 
whether or not it has decays to non-SM particles that escape undetected.  
Truly invisible decays include, for instance,  $H\to$~LSP+LSP (the LSP being the lightest supersymmetric 
particle in R-parity conserving supersymmetry and a DM candidate) 
while undetected, but not intrinsically invisible, decays are typified by $H\to AA$ where $A$ is a light pseudoscalar 
of an extended Higgs sector. In these and many other cases, ``invisible'' $H$ decays provide a portal to BSM physics 
that might prove hard to detect in any other way.  

In this paper, we have assessed the extent to which currently available data constrain invisible (or undetected) $H$ decays. By performing fits to all public data from the LHC and the Tevatron experiments, we have shown that the 95\% CL limits for $\brinv$ obtained depend very much upon the Higgs sector model.
Assuming a Higgs boson with SM couplings, $\brinv\simeq0.23$ is allowed at 95\%~CL.  
Allowing the $\cu$, $\cd$ and $\cv$ reduced coupling factors (that are defined as the coefficients multiplying 
the up-quark, down-quark and vector boson couplings relative to the SM values) to deviate from unity, 
much larger invisible decay rates are possible. In the most theoretically motivated case where $\cu>0$, $\cd>0$ and $\cv\leq 1$ ---the latter being required for any  two-doublet or two-doublet plus singlets model---we find $\brinv\leq 0.36$ at 95\%~CL.  

Limits on invisible (or yet undetected) Higgs decays also depend strongly on whether or not there are BSM particles that provide extra loop contributions to the $gg$ and $\gam\gam$ couplings of the $H$.  
In the simplest case of an SM-like Higgs with BSM loop contributions to its $gg$ and $\gam\gam$ couplings, the invisible decay rate can go up to about 60\% at 95\%~CL.  

In the absence of invisible decays, the best fits to the LHC data suggest significant deviations of $\cu$ and/or $\cp$ from unity.  We have shown that including the additional possibility of invisible/undetected $H$ decays makes even larger deviations accompanied by large $\brinv$ values consistent with LHC observations. 
If $\brinv$ really were as large as the $2\sigma$ limits we derived, one might hope that direct detection of invisible and/or hard-to-detect $H$ decays would be possible. Estimates suggest that ${\cal L}>300\fbi$ at $\rts=14\tev$ will typically be required.

Finally, we have also shown that if $\brinv\neq0$ is due to $H$ decays to a pair of DM particles, there are significant constraints on the size of $\brinv$ from the non-observation of spin-independent DM scattering on nucleons, the most important such limits currently being those from the XENON100 experiment.  
These constraints are much stronger for scalar DM than for Majorana or Dirac fermions. Overall, 
our results suggest a continued competition between limits on $\sigma_{\rm SI}$ and those on $\brinv$ as direct detection experiments achieve improved sensitivity and increasingly accurate measurements of the properties of the $H$ become available with future LHC running.

In short, precision measurements of the properties of the $H$ could well continue to provide the strongest constraints on a number of types of BSM physics, including the existence of light (mass $< m_H/2$) weakly interacting massive or hard-to-detect BSM particles.

\subsection*{Note added in proof} 

While this paper was being refereed, ATLAS and CMS presented major updates of their 
Higgs results based on $\sim25$~fb$^{-1}$ of data in most channels. Taking these new 
results into account,  we find 
$\brinv\leq0.20$  at 95\%~CL for a Higgs boson with SM couplings, and 
$\brinv\leq 0.29$ at 95\%~CL when  $\cu>0$, $\cd>0$ and $\cv\leq 1$ 
are left free to vary~\cite{inprep}.

\section*{Acknowledgements} 

GB thanks A. Pukhov for useful discussions. This work was supported in part by US DOE grant DE-FG03-91ER40674 and by IN2P3 under contract PICS FR--USA No.~5872. 
GB and SK acknowledge morever partial support from the French ANR  DMAstroLHC.
UE acknowledges partial support from the French ANR~LFV-CPV-LHC, ANR~STR-COSMO and the European Union FP7 ITN INVISIBLES (Marie Curie Actions,~PITN-GA-2011-289442). 

This work originated from discussions started at the Aspen Center for Physics (ACP) which is supported by the National Science Foundation Grant No.\ PHY-1066293; GB, UE, JFG and SK thank the ACP for hospitality and an inspiring working atmosphere. GB, JFG and SK also thank the Galileo Galilei Institute for Theoretical Physics (GGI Florence) for hospitality and the INFN for partial support. Finally, JFG acknowledges the hospitality of the Kavli Institute for Theoretical Physics which is supported
by the National Science Foundation under Grant No.  NSF PHY11-25915.



\providecommand{\href}[2]{#2}\begingroup\raggedright\endgroup

\end{document}